\documentclass[11pt,a4paper]{scrartcl}

\usepackage{CLICdp}
\usepackage{amsmath}
\usepackage{amssymb}
\usepackage{caption}    
\usepackage{subcaption}
\usepackage{csquotes}
\usepackage{comment}

\usepackage[backend=biber]{biblatex}

\usepackage{CLICdp_definitions}
\usepackage{CLICdp_biblatex}
\title{Lepton Flavor Violating Higgs decays at the Compact Linear Collider}

\clicdppub{2026}{001}  % journal articles
\date{27 April 2026}
\addauthor{Francisca Garay}{\institute{2}\hcomma\institute{3}\hcomma\editor{fmgaray@cern.ch}}
\addauthor{Gabriel Vega}{\institute{2}\hcomma\institute{3}}
\addauthor{Philipp Roloff}{\institute{1}}

\addinstitute{1}{CERN, Switzerland}
\addinstitute{2}{Pontificia Universidad Cat\'olica de Chile}
\addinstitute{3}{Millenium Institute for Subatomic Physics at the High Energy Frontier - SAPHIR}

\titlecomment{This work was carried out in the framework of the CLICdp Collaboration}

\abstract{
Lepton flavor violating decays of the Higgs boson $h$ could appear in scenarios beyond the Standard Model of particle physics. In this article, we study the sensitivity of a future Compact Linear Collider (CLIC), operating in $e^+e^-$ collisions, to such processes, namely, $h\rightarrow e\mu$, $h\rightarrow\tau\mu$, and $h\rightarrow e\tau$. In the absence of any observation, 95\% CL\ upper limits in the region of $10^{-4}$ to $10^{-5}$ could be placed on the branching fractions of these processes, assuming integrated luminosities of $4$\,ab$^{-1}$  at $\sqrt{s}=1.4$ TeV and $5$\,ab$^{-1}$ at $\sqrt{s}=3$ TeV. }

\graphicspath{ {./logos/}{./figures/} }

\begin{document}

% generates the title page
\titlepage

% include source for sections
%============================================%
% Guidelines and tips for CLICdp notes.
%
% Updated: 23.06.2014
% Christian Grefe (christian.grefe@cern.ch)
%============================================%

\newcommand{\latex}{\LaTeX\xspace}
\lstset{defaultdialect=[LaTeX]TeX}

\section{Introduction}
\label{sec:Intro}
One of the primary objectives in contemporary particle physics is to explore phenomena beyond the Standard Model (BSM). The discovery of a scalar Higgs boson $h$ at the LHC~\cite{ATLASHiggs,CMSHiggs} has offered significant insights into the mechanism of electroweak symmetry breaking~\cite{Higgs1,Higgs2,Higgs5}, thereby opening new avenues for investigating BSM physics within the Higgs sector.

The Standard Model (SM) of particle physics forbids lepton flavor violation (LFV) at  tree level. Such processes can only occur via loop-level contributions involving neutrinos, resulting in extremely small branching ratios. In particular, LFV Higgs decays, such as $h \rightarrow \tau\mu$, $h \rightarrow \tau e$, and $h \rightarrow e\mu$, are suppressed by the tiny neutrino masses, rendering them far below any conceivable experimental sensitivity. Consequently, the observation of any Higgs boson decays into two leptons with different flavors would be a clear indication of new physics.

While neutrino oscillations demonstrate that lepton flavor is not strictly conserved,  there is no LFV at a single interaction vertex. The observed violation in the neutrino sector hints that other sources of LFV could exist, including in the charged lepton sector. Such decays naturally arise in various extensions of the SM, including models with multiple Higgs doublets~\cite{LFVTheo1,LFVTheo2,LFVTheo3,LFVTheo4,LFVTheo5}, composite Higgs models~\cite{LFVTheo6,LFVTheo7}, models with flavor symmetries~\cite{LFVTheo8}, warped extra dimensions~\cite{LFVTheo7,LFVTheo9,LFVTheo10}, and other theoretical frameworks~\cite{LFVTheo11,LFVTheo12}. Previous hints of deviations from lepton universality in rare $B$-meson decays observed by BaBar and LHCb~\cite{Babar,LHCb1,LHCb2} have motivated models of new physics that also predict lepton flavor violating (LFV) processes, including LFV decays of the Higgs boson~\cite{pheno1,pheno2,pheno3}. While recent LHCb measurements~\cite{LHCbRK_PRL2023,LHCbRKstar_PRD2023} show consistency with SM expectations, the theoretical interest in LFV signatures remains strong, particularly given the enhanced sensitivity of future lepton colliders such as CLIC,which is designed to operate in $e^+e^-$ collisions at centre-of-mass energies ranging from a few hundred GeV to multi-TeV scales.

The ATLAS and CMS experiments at the Large Hadron Collider (LHC) have searched for LFV Higgs decays. The most stringent 95\% confidence level (CL) limits are $BR(h \rightarrow \tau\mu) < 0.0018$ (ATLAS) and 0.0015 (CMS), and $BR(h \rightarrow \tau e) < 0.002$ (ATLAS) and 0.0022 (CMS)~\cite{ATLASLFV, CMSLFV}. The $h \rightarrow e\mu$ decay has been probed only by ATLAS, with $BR(h \rightarrow e\mu) < 6.1 \times 10^{-5}$ at 95\% CL~\cite{ATLASLFV2}. All results are based on $\sqrt{s} = 13$~TeV $pp$ collisions corresponding to an integrated luminosity of about $139~\mathrm{fb}^{-1}$.

Future lepton colliders provide an excellent environment to search for LFV Higgs boson decays. An earlier phenomenological study evaluated the sensitivity to the $h\rightarrow e\mu$, $h\rightarrow\tau\mu$, and $h\rightarrow e\tau$ channels at the CEPC, FCC-ee, and ILC~\cite{Futureee}. More recently, the projected sensitivity of the FCC-ee to the same channels at $\sqrt{s}=240$~GeV was investigated~\cite{FCCeeLFV}, providing a complementary assessment to the CLIC projections presented in this work.

The most stringent indirect constraint on the $h \rightarrow e\mu$ decay arises from searches for $\mu \rightarrow e\gamma$ decays~\cite{pheno4}, yielding a bound of $BR(h \rightarrow e\mu) \leq O(10^{-8})$~\cite{pheno5, pheno6}.

In this study, we investigate the potential sensitivity of the CLIC facility to LFV Higgs boson decays at both $\sqrt{s}=1.4$~TeV and $\sqrt{s}=3$~TeV energy stages. The initial CLIC operation at $\sqrt{s}=380$~GeV is not considered because the Higgs production rate at this energy, dominated by the Higgsstrahlung process ($e^+e^-\rightarrow ZH$), is significantly smaller than at the higher-energy stages, where Higgs production via $WW$ fusion ($e^+e^-\rightarrow h\nu\bar{\nu}$) dominates, as illustrated in Figure~1. Consequently, the expected sensitivity to these rare decays is substantially reduced at $\sqrt{s}=380$~GeV.

%In this study, we investigate the potential sensitivity of the CLIC facility to LFV Higgs decays at both $\sqrt{s} = 1.4$~TeV and $\sqrt{s} = 3$~TeV energy stages. In both cases, the dominant production process for a SM Higgs boson is $e^+e^- \rightarrow h\nu\bar{\nu}$, as illustrated in Figure~\ref{fig:cs}.

\begin{figure}[tbh]
    \centering
    \includegraphics[scale=0.3]{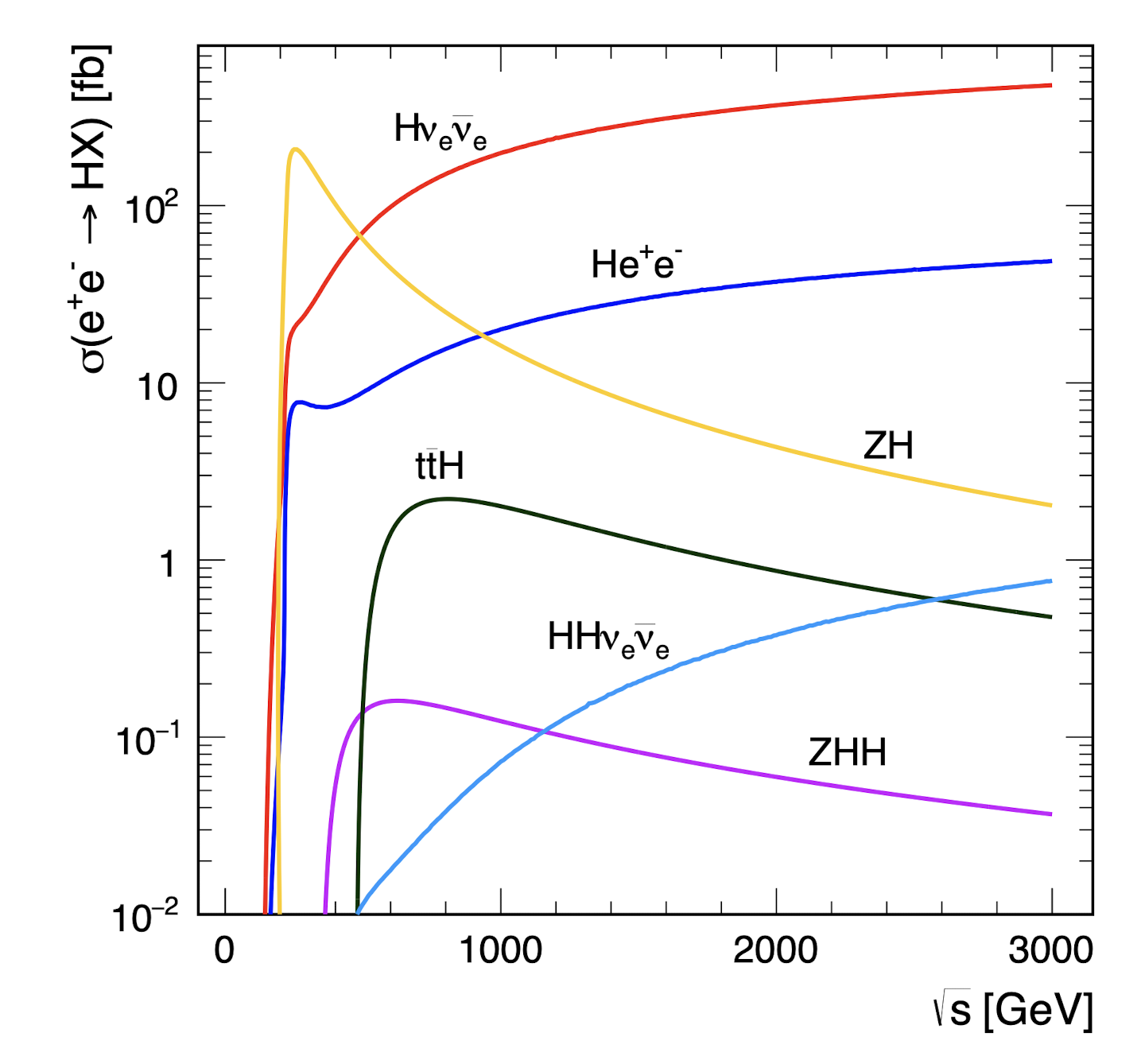} 
    \caption{ Cross section as a function of center-of-mass energy for the main Higgs production processes at an $e^+e^-$ collider for a Higgs mass of $m_h=126$ GeV~\cite{CLICHIGGS}. The values shown correspond to unpolarised beams and do not include the effect of beamstrahlung.}
    \label{fig:cs}
\end{figure}

\section{The Compact Linear Collider CLIC}
\label{sec:CLIC}
To maximize its physics potential, the CLIC accelerator technology itself presents conditions that are challenging for an experiment. The detector concepts required for the precise reconstruction of complex final states are designed to address these and to operate across a wide range of collision energies. The accelerator will provide high-luminosity collisions at center-of-mass energies of 380~GeV, 1.4~TeV and 3~TeV~\cite{CLIC_1} with corresponding integrated luminosities of 1~ab$^{-1}$, 4~ab$^{-1}$ and 5~ab$^{-1}$, respectively~\cite{CLIC_2,CLIC_3}. 

The accelerator is based on an innovative two-beam acceleration scheme, in which normal-conducting high-gradient 12~GHz X-band accelerating structures are powered via a high-current drive beam~\cite{CLIC_4}. This enables a compact and cost-effective accelerator complex, with a site length ranging between 11 and 50~km.

The accelerator complex is accompanied by a multipurpose detector system. The CLIC\_ILD detector concept, used for the study described here, is adaptated from the ILD detector concept~\cite{ILC_2} that was developed for the International Linear Collider (ILC).

The inner part of the CLIC\_ILD detector consists of a central gaseous time projection chamber (TPC) for tracking, enclosing an ultra-thin silicon-pixel vertex detector. In addition, the TPC is surrounded by a silicon strip detector envelope. These provide excellent charged particle momentum resolution and impact parameter resolution, crucial for a precise vertex reconstruction and flavor tagging.

\begin{figure}[tbh]
    \centering
    \includegraphics[scale=0.5]{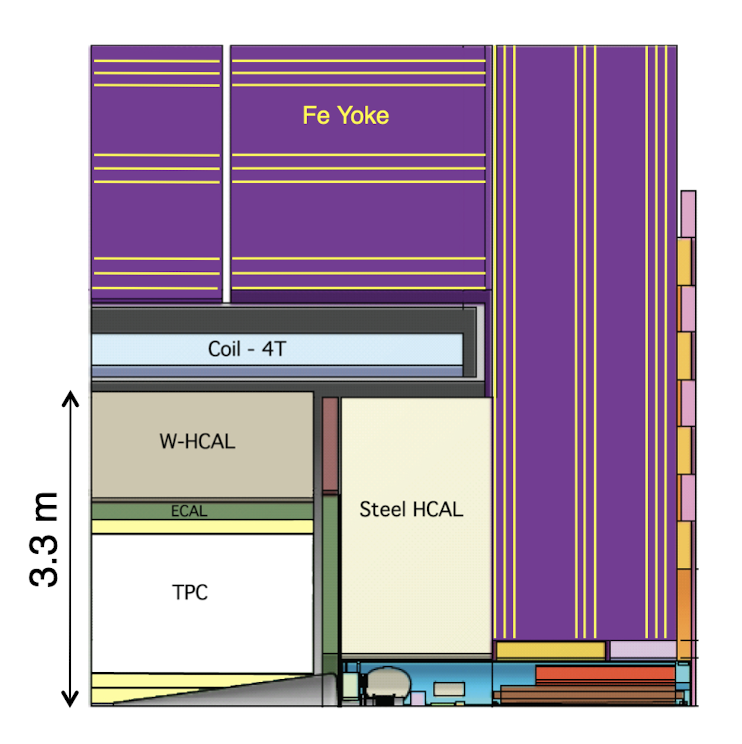} 
    \caption{ Longitudinal cross section of the top right quadrant of the CLIC\_ILD detector concept~\cite{Linssen:2012hp}.}
    \label{fig:clic_ild}
\end{figure}

The vertex and tracking systems are surrounded by a highly-granular sampling calorimetry system, composed of an electromagnetic and a hadronic calorimeter (ECAL and HCAL), optimized for particle flow reconstruction. The expected jet-energy resolution, for isolated central light-quarks jets with energy in the range from 100~GeV to 1~TeV, is $\sigma_E/E\lesssim 3.5\%$~\cite{CLIC_5}. A strong solenoidal magnet located outside the HCAL provides a 4~T magnetic field. The magnetic flux return is contained in a large iron yoke instrumented with detectors for muons identification. Dedicated calorimeters in the very forward region are used for luminosity measurements and extended coverage for electrons and photons. The detector layout is shown in Figure~\ref{fig:clic_ild},  for details see~\cite{Linssen:2012hp}.
 
The beam and bunch structure of CLIC consists of bunch trains of 312 bunches, separated from each other by 0.5~ns, with a bunch train repetition frequency of 50~Hz. High luminosity is achieved by ultra-low beam emittance and hence  highly-focused and intense electron and positron beams at the interaction point. This generates significant beamstrahlung radiation between colliding bunches~\cite{Linssen:2012hp}, an effect that is considered in this study.

\section{Event Simulation, Background Processes and Reconstruction}
\label{MCGen}

The signal and main physics background processes were generated using the WHIZARD 1.95~\cite{Whiz} package. The expected energy spectra of the CLIC beams for $\sqrt{s} = 1.4$ and 3~TeV were used for the initial-state electrons and positrons. It was found in previous studies of the $h \to \mu^{+}\mu^{-}$ and $h \to \tau^{+}\tau^{-}$ final states~\cite{Linssen:2012hp,CLICHIGGS} that backgrounds from $e^{+}e^{-}$ interactions represent the largest backgrounds after typical event selections. For this reason, processes with hard photons in the initial state were neglected in the studies presented in this paper. Initial-state radiation (ISR) was parameterised by the leading logarithmic approximation structure function including hard collinear photons up to the third order~\cite{Skrzypek:1990qs}. For consistency with other CLIC physics studies, a Higgs boson mass of 126~GeV was assumed \cite{CLICHIGGS,CLIC_1,CLIC_2}. The fragmentation and hadronization steps were simulated using PYTHIA 6.4~\cite{Sjostrand:2006za} using parameters tuned to OPAL data~\cite{OPAL:1995cgp} as described in~\cite{Linssen:2012hp}. The Higgs boson decays were also generated using PYTHIA and have been added to its decay table. Finally, tau-lepton decays were generated with the TAUOLA package~\cite{Was:2000st}. 

The samples used in this study assume unpolarized beams. The current CLIC baseline~\cite{Robson:2025} assumes 80\% of data taken with -80\% electron polarization and 20\% of data taken with +20\% electron polarization. The overall enhancement of single Higgs production in $WW$ fusion is 1.48 with respect to the unpolarized case (compared to 1.8 for a pure -80\% electron polarization~\cite{CLICHIGGS}). A similar improvement may be expected for the LFV signal, although a quantitative estimate would require polarized signal and background samples.

The detector simulation and event reconstruction chain was carried out using the CLIC\_ILD detector model~\cite{Linssen:2012hp}. The performance of this CLIC\_ILD detector is very similar to the newer model, for details see~\cite{CLICdet_2}. Pile-up from $\gamma\gamma\rightarrow$ hadrons interactions was overlaid to the physics events. Event reconstruction is performed using the MARLIN software package~\cite{Gaede:2006pj}, while Pandora PFA~\cite{CLIC_6,CLIC_7} is used for calorimeter clustering and particle flow reconstruction, creating a collection of so-called Particle-Flow Objects (PFOs). The signal samples correspond to the Higgs boson production process
$ee \rightarrow h\nu\nu$,
followed by the LFV decays
$h\rightarrow e\mu$,
$h\rightarrow\tau\mu$, and
$h\rightarrow e\tau$.
The dominant background processes are electroweak four-fermion production,
$ee \rightarrow \ell\ell'\nu\nu$, where $\ell,\ell'=e,\mu,\tau$, including both same- and different-flavour charged-lepton final states, and hadronic production,
$ee \rightarrow qq\nu\nu$.
The latter contributes primarily to the channels containing a reconstructed $\tau$ lepton through the misidentification of hadronic jets as $\tau$-lepton candidates. The impact of this background is quantified by the TauFinder fake-rate study presented in Section~4.2.1. The signal and background samples considered in this study are summarized in Table~1.

%A summary of the simulated samples used in this study is given in Table~\ref{Tab:mcsamples}, \textcolor{red}{where the signal cross sections are computed assuming $BR(h\rightarrow ll^{\prime})=1$.} \textcolor{red}{The dominant background processes are four-fermion final states $ee\rightarrow \ell\ell^{\prime}\nu\nu$, where $l,l^{\prime}=e,\mu,\tau$ including all flavour combinations relevant to the signal channels.}

\begin{table}[tbh]
\centering
\begin{tabular}{| c|c|c |} 
 \hline
 Process & \multicolumn{2}{c|}{$\sigma[\mathrm{fb}]$} \\  \cline{2-3}
  & $\sqrt{s}=1.4$~TeV & $\sqrt{s}=3.0$~TeV\\
  \hline
 $ee\rightarrow h\nu\nu,\, h\rightarrow e\mu$ & 244.0 &  415.1 \\ 
  \hline
 $ee\rightarrow h\nu\nu,\, h\rightarrow e\tau$ & 244.0 &  415.1 \\ 
  \hline
  $ee\rightarrow h\nu\nu,\, h\rightarrow \mu\tau$ & 244.0 &  415.1  \\ 
   \hline
$ee\rightarrow \ell\ell^{\prime}\nu\nu$ & 978.5 &  946.5 \\ 
 \hline
 $ee\rightarrow qq\nu\nu$ & 787.7 & 1318 \\ 
 \hline
\end{tabular}
 \caption{Signal and background processes considered. The signal cross sections are computed assuming $BR(h\rightarrow ll^{\prime})=1$, where $q$ represents any quark other than $t$ and $l,l^{\prime}$ can be any pair of lepton species, where the minimum  mass of charged lepton pairs generated is 50~GeV and $m_h=126$~GeV. }
 \label{Tab:mcsamples}
\end{table}
A right-handed coordinate system is used, with the origin at the nominal interaction point. The $z$-axis is defined along the electron beam direction, the $x$-axis points horizontally, and the $y$-axis points vertically upward. The polar angle $\theta$ is measured from the $+z$-axis, and the azimuthal angle $\phi$ is measured around the beam axis from the $+x$-axis. This convention corresponds to the coordinate system used in the detector simulation and analysis.

\section{Lepton Flavor Violating Higgs decays}

We investigated the sensitivity of CLIC to LFV Higgs decays, focusing on the three possible decay channels: \( h \rightarrow e\mu \) and two inclusive $\tau$ leptonic modes \( h \rightarrow \tau\mu \) and \( h \rightarrow e\tau \).
%Additionally, we analyzed these decays at two different energy stages: \( \sqrt{s} = 1.4 \, \text{TeV} \) and \( \sqrt{s} = 3 \, \text{TeV} \).
%
The variables used for the analyses are defined as follows:
%of the different $h\rightarrow \ell\ell^{\prime}$ channels are:

\begin{itemize}
    \item $m_{\ell\ell^\prime}$: The invariant mass of the dilepton system. In final states containing electrons, Bremsstrahlung photons were incorporated into the invariant mass calculation.
    \item $|\Delta \theta_{\ell\ell^\prime}|$: The absolute value of the difference between the polar angles ($\theta$) of the leptons.
    \item $|\Delta \phi_{\ell\ell^\prime}|$: The absolute value of the difference between the azimuthal angles, $\phi$ of the leptons.
    \item $\cos\theta^*$: The cosine of the helicity angle\footnote{The helicity angle $\theta^*$ is defined as the angle between the momentum of one lepton in the dilepton rest frame and the boost direction of the dilepton system.} of the dilepton system.
    \item $\beta_{\ell\ell^\prime}$: The boost of the dilepton system in the lab frame. 
    \item $E_{\text{vis}}$: The visible energy in the event.
    \item $E_{\text{rest}}$: The sum of the energies of the event without the dilepton system.
    \item $\theta_{\text{miss}}$: The polar angle of the missing momentum vector.
\end{itemize}

\subsection{\texorpdfstring{$h\rightarrow e\mu$}{h->e mu} decay channel}

% Measuring the LFV Higgs boson decay into an electron and a muon in the final state presents the greatest challenge among the three possible decays due to the electron's small mass, resulting in the lowest branching ratio of the three channels. 
This final state is the cleanest, as both final state particles are stable. In the \( ee \rightarrow h \nu \nu \) production process, the \( h \rightarrow e\mu \) decay is identified by an electron-muon pair with opposite charges that is consistent with the Higgs boson mass and accompanied by missing momentum. The excellent momentum resolution of the detector enables high background rejection and signal selection efficiency. Signal and background events were reconstructed as described in Section~\ref{MCGen}.
The primary background processes are \( ee \rightarrow \ell\ell^{\prime}\nu\nu \) with $\ell=e$ or $\mu$ , as listed in Table~\ref{Tab:mcsamples}.

\begin{figure}[tbh]
     \centering
     \begin{subfigure}[b]{0.47\textwidth}
         \centering
         \includegraphics[width=\textwidth]{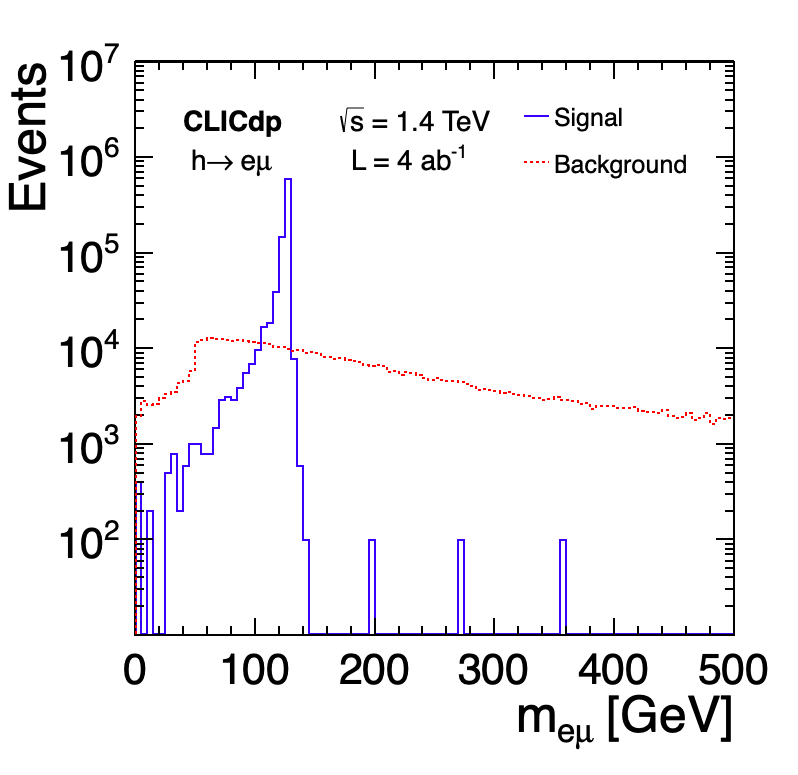}
         \caption{}
         \label{m_emu}
     \end{subfigure}
     \hfill
     \begin{subfigure}[b]{0.47\textwidth}
         \centering
         \includegraphics[width=\textwidth]{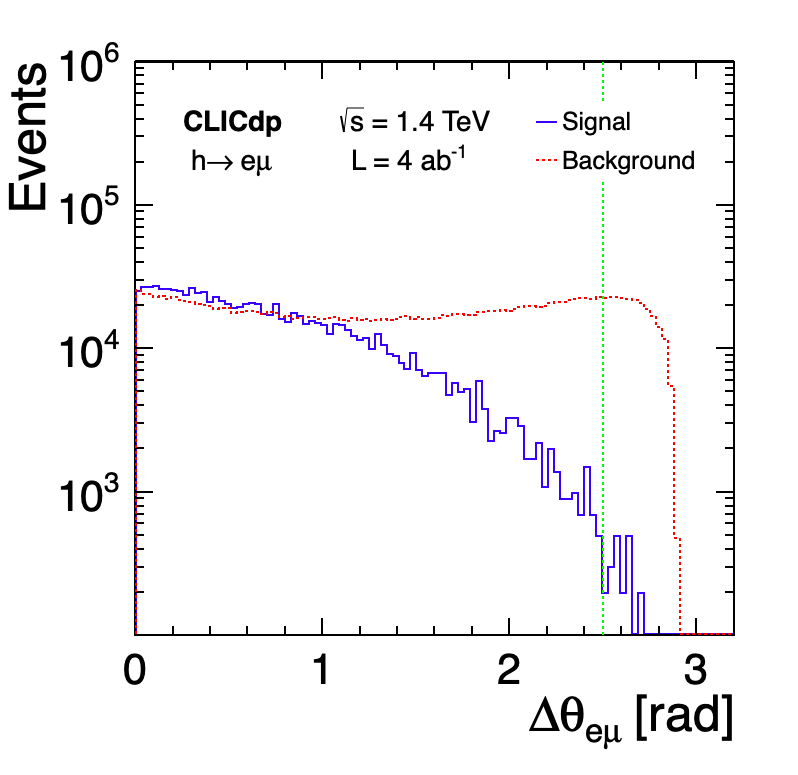}
         \caption{}
         \label{delta_theta_emu}
     \end{subfigure}
     \hfill
        \caption{(a) Invariant mass distribution and (b) the difference between the angles of the particles of the dilepton system $e\mu$ for signal and background at $1.4$~TeV and assuming integrated luminosity of $\mathcal{L}=4$~\abinv. The green line represents the cut applied on the variable.}
        \label{fig_emu1}
\end{figure}

\begin{figure}[tbh]
     \centering
     \begin{subfigure}[b]{0.47\textwidth}
         \centering
         \includegraphics[width=\textwidth]{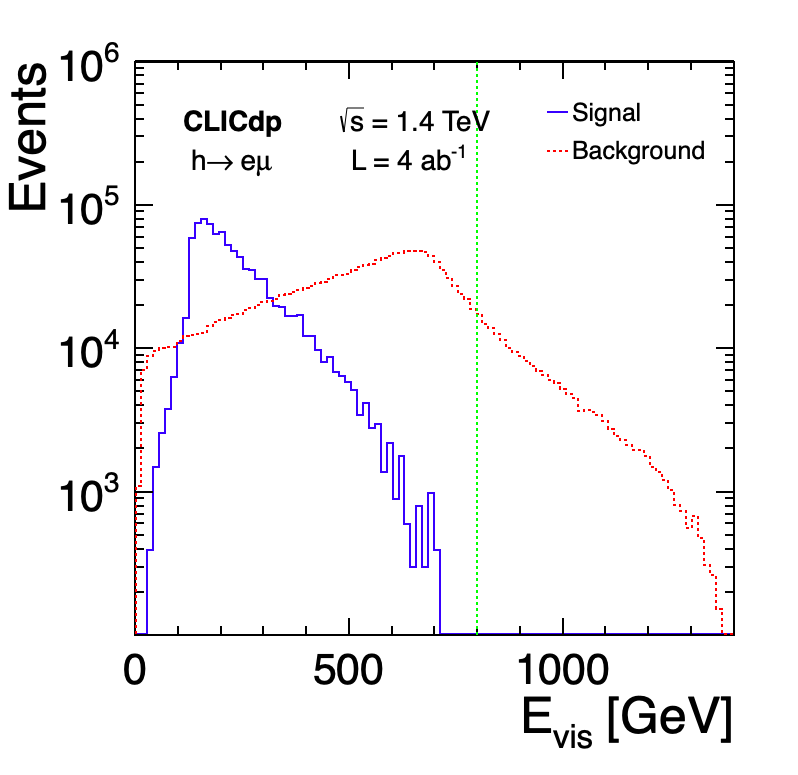}
         \caption{}
         \label{E_vis}
     \end{subfigure}
     \hfill
     \begin{subfigure}[b]{0.47\textwidth}
         \centering
         \includegraphics[width=\textwidth]{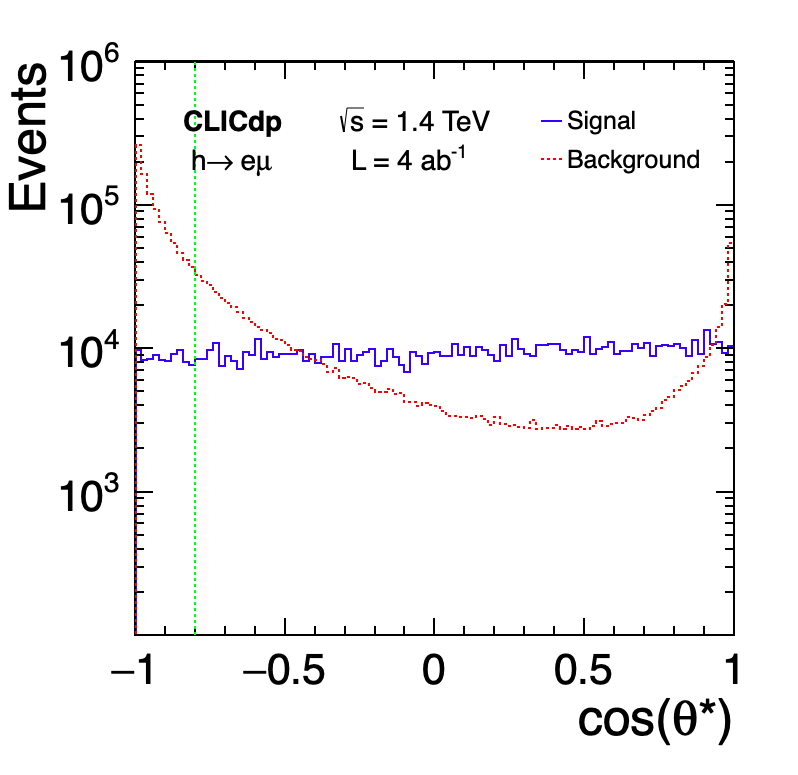}
         \caption{}
         \label{costheta_emu}
     \end{subfigure}
     \hfill
        \caption{(a) Visible energy distribution of the events and (b) the cosine of the helicity angle of the dilepton system $e\mu$ for signal and background at 1.4~TeV and assuming integrated luminosity of $\mathcal{L}=4$~\abinv. The green line represents the cut applied on the variable.}
        \label{fig_emu2}
\end{figure}

The event selection criteria require two reconstructed leptons of opposite charge and different flavors ($e$, $\mu$). Additionally, an electron energy threshold of $E_e > 8$ GeV was applied to avoid poorly reconstructed electrons, as those below this energy fall outside the detector's acceptance. For $1.4$ TeV collisions, the cuts $|\Delta\theta_{e\mu}| < 2.5\, \text{rad}$, $\cos\theta^*>-0.8$ and $E_{\text{vis}}<800$ GeV were applied. As an example, Figures~\ref{fig_emu1} and~\ref{fig_emu2} show four representative distributions for this channel at $\sqrt{s}=1.4~\mathrm{TeV}$. In Figure~\ref{fig_emu2}, the vertical green line indicates the value of the preselection cut applied in the analysis.
Figure~\ref{m_emu} compares  the invariant mass distribution of the dilepton system. The small tail observed on the left side of this distribution originates mainly from Bremsstrahlung photons that could not be fully recovered. In Figure~\ref{delta_theta_emu}, we show the distribution of the opening-angle difference $\Delta\theta_{e\mu}$ between the leptons of the system. 
The background exhibits the expected behaviour for multi-body electroweak processes, with a large population at small angular separations and a steep fall-off towards $\Delta\theta \simeq \pi$. Signal events, coming from a two-body Higgs decay, also populate a wide range of angles but remain subdominant across the full spectrum, as expected. Figure~\ref{E_vis} shows the distribution of the visible energy 
$E_{\mathrm{vis}}$ of the events. Signal events exhibit a peak at moderate visible energies, consistent with the kinematics of $h\to e\mu$ decays at this center-of-mass energy. Background events, in contrast, populate a much broader range and develop a long high-energy tail, also as expected. Figure~\ref{costheta_emu} shows the distribution of the cosine of the helicity angle of the dilepton system, $\cos\theta^*$. 
Signal events, originating from the decay of a Higgs boson, display 
a nearly flat distribution, consistent with an isotropic decay in the Higgs rest frame. Background events feature a strong enhancement at large negative values of $\cos\theta^*$, characteristic of t-channel dominated electroweak processes. Both behaviors are consistent with expectations for signal and background.

\begin{figure}[tb]
     \centering
     \begin{subfigure}[b]{0.47\textwidth}
         \centering
         \includegraphics[width=\textwidth]{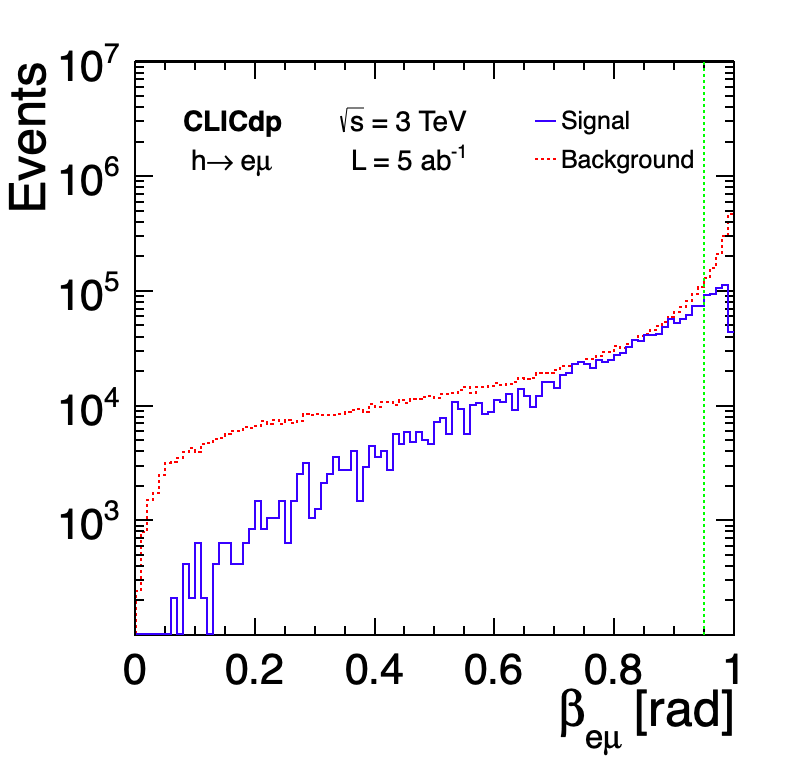}
         \caption{}
         \label{beta_higgs_3TeV}
     \end{subfigure}
     \hfill
     \begin{subfigure}[b]{0.47\textwidth}
         \centering
         \includegraphics[width=\textwidth]{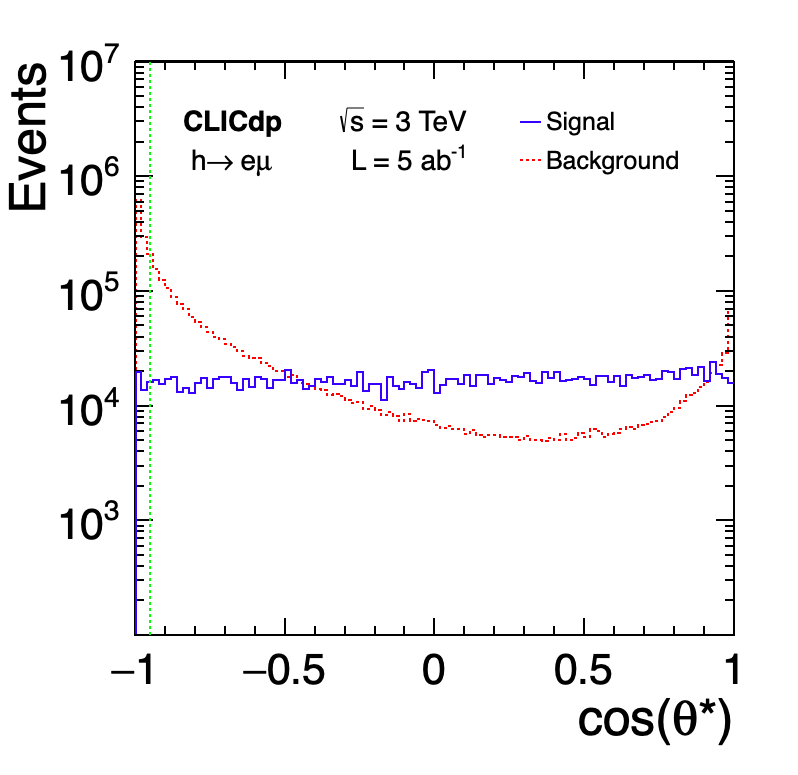}
         \caption{}
         \label{costheta_emu_3Tev}
     \end{subfigure}
     \hfill
        \caption{(a) The boost of the dilepton system in the lab frame and (b) the cosine of the helicity angle of the $e\mu$ dilepton system for signal and background at $3~\text{TeV}$, assuming an integrated luminosity of $\mathcal{L}=5~\text{ab}^{-1}$. The green line represents the cut applied on the variable.}
        \label{fig_emu3}
\end{figure}

For 3~TeV collisions, the cuts $\beta_{\text{e}\mu}<0.95$ and $\cos\theta^*>-0.95$ were used. In Figure~\ref{beta_higgs_3TeV}, we show the boost distribution of the $e\mu$ dilepton system in the laboratory frame, $\beta_{e\mu}$. The background shows a more pronounced rise at small and large $\beta_{e\mu}$, while the signal tends to higher values in this region. In Figure~\ref{costheta_emu_3Tev}, we show the distribution of the helicity angle of the dilepton system, expressed as $\cos\theta^*$, which displays the same qualitative behaviour as at $\sqrt{s}=1.4~\text{TeV}$.

This set of preselection cuts was applied to suppress the dominant backgrounds, designed to reject events with poorly reconstructed objects or soft leptons, while maintaining high signal efficiency. At 1.4~TeV, the most effective cut is over $\cos\theta^*$, which removes over half of the background while retaining more than 90\% of the signal, since extreme backward configurations (where one lepton has low momentum/energy in the laboratory frame) are more common in background. Additional requirements on $\Delta\theta_{e\mu}$ and $E_{\text{vis}}$, provide further suppression of widely separated or high-activity events, with negligible impact on the signal.
At 3~TeV, the $\beta_{e\mu}$ cut is most powerful, removing about 42\% of the background (at some signal cost) since the Higgs will be highly boosted, while $\cos\theta^*$ selection further suppresses extreme backward configurations. After all cuts, $87.4\%$ of the signal events remain, and the background is reduced to 15.1\% at 1.4~TeV, while 57.9\% of the signal events remain, and the background is reduced to 7.0\% for the $3$ TeV sample. An overview of the selection cut results is summarized in Table~\ref{Tab:cutsemu} and Table~\ref{Tab:cutsemu_3TeV}.

\begin{table}[tbh]
\centering
\begin{tabular}{ |c|c|c|c|} 
 \hline
 Process & Events & Efficiency  & Exp. Events\\ 
 \hline
$ee\rightarrow h\nu\nu\, (h\rightarrow e\mu)$ & $9.9$k & 95.8\% & $935$k $\pm 9$k \\ 

\hline
$ee\rightarrow \ell\ell^{\prime}\nu\nu$ & $1.57$M & 41.7\% & $1.67$M $\pm 2$k\\ 

\hline
\end{tabular}
\caption{Number of simulated signal and background events surviving the event selection (``Events'') at $\sqrt{s}=1.4$~TeV, together with the corresponding expected event yields (``Exp. Events'') for an integrated luminosity of $\mathcal{L}=4~\mathrm{ab}^{-1}$and assuming a branching ratio of 1. The quoted uncertainties are statistical only and originate from the finite size of the simulated samples.}
 %\caption{Number of events in the signal and background samples at 1.4 TeV. The expected signal yield is computed assuming a branching ratio of 1. All numbers are normalized to an integrated luminosity of $\mathcal{L} = 4\,\mathrm{ab}^{-1}$.} 
 \label{Tab:cutsemu}
\end{table}

\begin{table}[tbh]

\centering
\begin{tabular}{|c|c|c|c|} 
\hline
 Process & Events & Efficiency & Exp. Events\\ 
\hline
$ee\rightarrow h\nu\nu\, (h\rightarrow e\mu)$ & $9.9$k  & 40.1\% & $832$k $\pm 2$k\\ 

$ee\rightarrow \ell\ell^{\prime}\nu\nu$  & $1.55$M & 34.9\% & $1.68$M $\pm 18$k\\
\hline
\end{tabular}
 \caption{Number of events in the signal and background samples at 3 TeV. The expected signal yield is computed assuming a branching ratio of 1. All numbers are normalized to an integrated luminosity of $\mathcal{L} = 5\,\mathrm{ab}^{-1}$. The notation is the same as in Table \ref{Tab:cutsemu}.\label{Tab:cutsemu_3TeV}}
\end{table}

\begin{figure}[tb]
     \centering
     \begin{subfigure}[b]{0.49\textwidth}
         \centering
         \includegraphics[width=\textwidth]{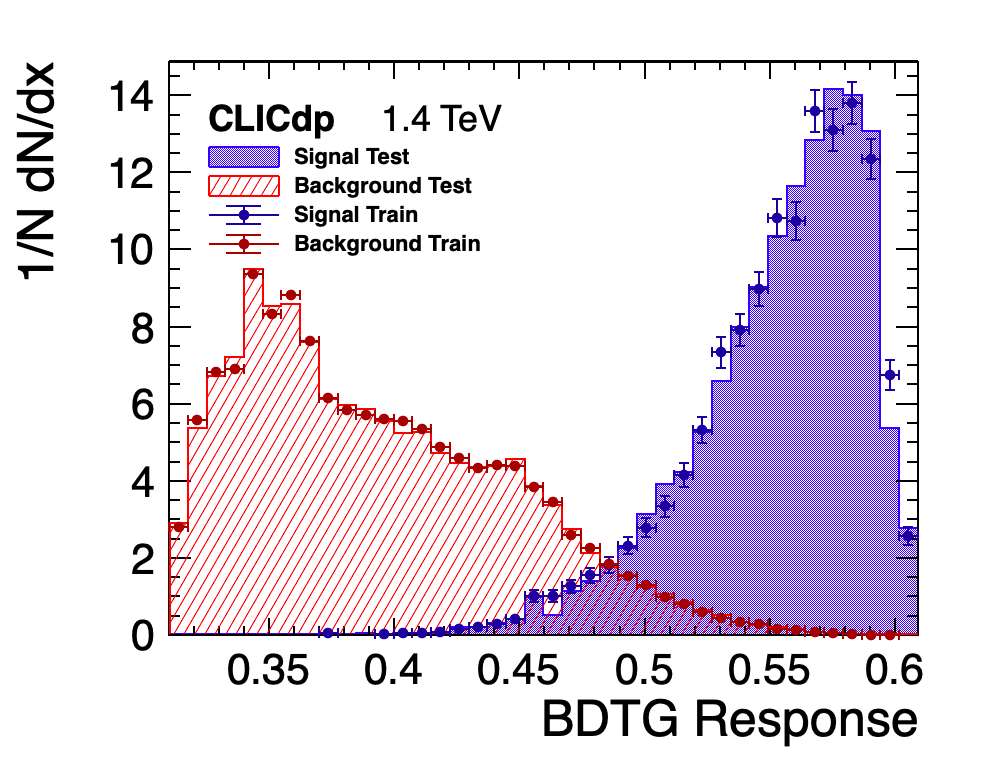}
         \caption{}
         \label{BDT_emu}
     \end{subfigure}
     \hfill
     \begin{subfigure}[b]{0.5\textwidth}
         \centering
         \includegraphics[width=\textwidth]{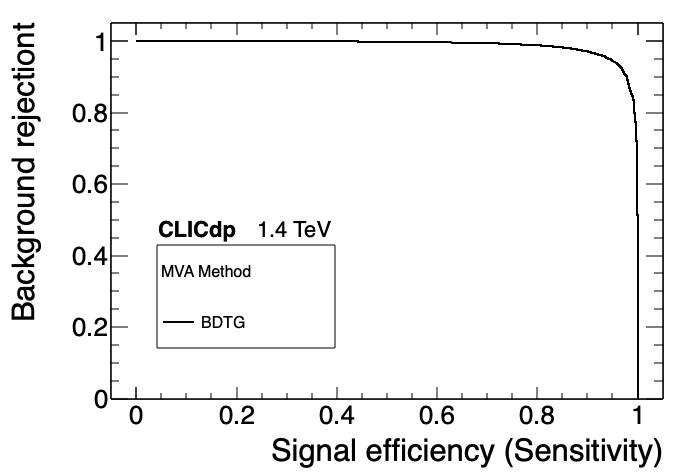}
         \caption{}
         \label{fig:ROC}
     \end{subfigure}
     \hfill
        \caption{(a) BDT response distribution for signal and background samples for the $h\rightarrow e \mu$ channel at 1.4 TeV. The Kolmogorov-Smirnov test results for the signal and background are 0.957 and 0.446 respectively and (b) ROC Curve showing signal efficiency vs. background rejection for the BDT at 1.4~TeV. The area under the curve has a value of 0.989.}
\end{figure}

The final event selection employs a Boosted Decision Tree (BDT) classifier~\cite{TMVA:2007ngy} using a subset of the kinematic variables introduced previously. Several kinematic observables, including $\Delta\theta_{e\mu}$, $\cos\theta^{*}$, and $\beta_{e\mu}$, are used both in the preselection and as input variables to the BDT. The preselection applies relatively loose requirements to reject events in background-dominated regions of phase space while maintaining a high signal efficiency. The BDT subsequently exploits the residual discriminating power of these variables, together with their correlations with the other kinematic observables, to achieve optimal signal-background separation. At 1.4~TeV, those used are: $m_{e\mu}$, $|\Delta\theta_{e\mu}|$, $|\Delta\phi_{e\mu}|$, $\beta_{e\mu}$, $\cos\theta^*$ and $E_{\text{vis}}$.  At 3~TeV, the selected variables are: $m_{e\mu}$, $|\Delta\theta_{e\mu}|$, $|\Delta\phi_{e\mu}|$, $\beta_{e\mu}$, $\cos\theta^*$, $E_{\text{rest}}$ and $\theta_{\text{miss}}$. In selecting these variables, we examined the correlation matrices and ensured that no pair exhibited a correlation greater than 73\% in either the signal or background samples, for the two energy stages. In addition, the Higgs boson mass will be precisely determined from other channels at the HL-LHC and at CLIC, allowing the signal MC samples to be tuned reliably. Consequently, the correlation between the dilepton invariant mass and the BDT response does not affect the robustness of this analysis. The resulting BDT output distribution and the Receiver Operating Curve (ROC) are shown in Figures~\ref{BDT_emu} and \ref{fig:ROC} at 1.4~TeV, as an example. High background rejection and good signal efficiency are obtained for both energy schemes, since the area under the ROC curve has a value of 0.989 at 1.4~TeV and 0.994 at 3~TeV.

The BDT score distribution is compared with the background-only BDT score distribution to derive the expected limit on $BR(h \rightarrow e\mu)$. Applying the asymptotic approximation of the $CL_s$ method~\cite{stats}, the expected 95\% CL upper limit, assuming no signal is observed, for an integrated luminosity of 4~ab$^{-1}$ at $\sqrt{s} = 1.4$~TeV without electron beam polarization, is
$$BR(h\rightarrow e\mu)<6.22\times 10^{-5}.$$
Similarly, the equivalent expected 95\% CL limit  at a luminosity of $5\,\mathrm{ab}^{-1}$ at $\sqrt{s}=3$~TeV, is 
$$BR(h\rightarrow e\mu)<1.42\times 10^{-5}.$$
Background normalization uncertainties have a negligible impact on the expected limits, with changes below \(0.04\%\) for uncertainties up to 5\% at both energy stages. The corresponding expected \(5\sigma\) discovery reach is \(\mathrm{BR}(H\to e\mu)=2.24\times10^{-4}\) at 1.4 TeV and \(1.51\times10^{-4}\) at 3 TeV.

\begin{figure}[tbh]
    \centering
    \includegraphics[width=0.6\linewidth]{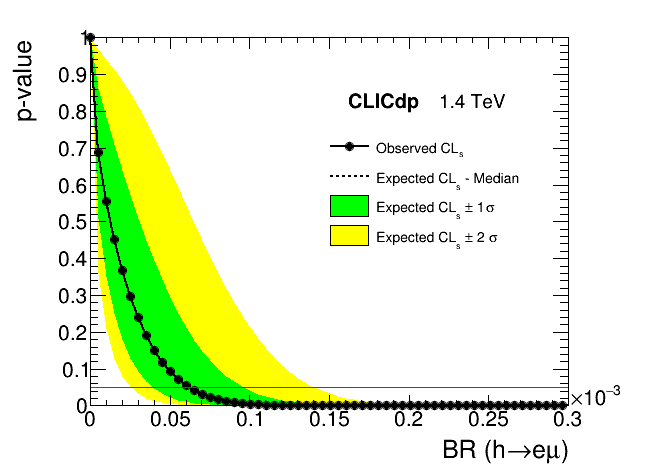}
    \caption{Asymptotic $CL_s$ scan for the branching ratio $BR(h\rightarrow e\mu)$ at $\sqrt{s}=1.4$~TeV. The green and yellow bands represent the expected $\pm1\sigma$ and $\pm2\sigma$ variations, respectively, around the median expected $CL_s$ under the background-only hypothesis.}
    %\caption{Asymptotic $CL_s$ scan for the branching ratio $BR(h\rightarrow e \mu)$ at 1.4~TeV.}
    \label{cls_emu}
\end{figure}

Figure~\ref{cls_emu} presents the CLs scan for various branching ratio values at 1.4~TeV, as an example. The $p$-values shown  correspond to the values obtained for different assumed branching ratios at $\sqrt{s}=1.4~\mathrm{TeV}$ using the $\mathrm{CL}_\mathrm{s}$ scan method. For a given assumed signal branching ratio, it quantifies the level of compatibility of the observed data with the signal-plus-background hypothesis relative to the background-only hypothesis. Smaller $p$-values therefore indicate a stronger exclusion of the corresponding signal hypothesis. The red horizontal line indicates the $p$-value threshold corresponding to $\mathrm{CL}_\mathrm{s}=0.05$, which defines the 95\% CL exclusion. The intersection of the observed (expected) $\mathrm{CL}_\mathrm{s}$ curve with this line determines the observed (expected) upper limit on the branching ratio. These results indicate that CLIC at 1.4~TeV could achieve similar sensitivity than reported by the ATLAS experiment \cite{ATLASLFV2}, increasing to a factor of 4.5 times at 3~TeV\footnote{The ATLAS and CMS limits are obtained from a full experimental analysis including detailed background modeling and systematic uncertainties \cite{ATLASLFV,ATLASLFV2,CMSLFV}, whereas our CLIC study is phenomenological and does not incorporate the full range of detector-related effects. The comparison, in all the channels, should thus be regarded as indicative rather than directly comparable.}.

\subsection{\texorpdfstring{$h\rightarrow\tau l$}{h->tau l} decays and \texorpdfstring{$\tau$}{tau} reconstruction}

We will present results for both channels where  $l$ represents either electrons or muons. To proceed, we will first provide a description of the $\tau$-lepton reconstruction process.

\subsubsection{\texorpdfstring{$\tau$}{tau}-lepton reconstruction using TauFinder}

The reconstruction of $\tau$ leptons is carried out using the TauFinder reconstruction tool~\cite{taufinder}, which employs a seeded cone-based jet clustering algorithm. The algorithm reconstructs hadronically decaying $\tau$-lepton candidates from their visible decay products, providing the four-momentum (energy and momentum) and electric charge of each $\tau$-lepton candidate. These reconstructed quantities are used throughout the subsequent event selection. This approach depends on specific steering parameters, including the opening angles for the search and isolation cones, and has been optimized for the $\tau$ lepton signature relevant to our analysis. In this study, a parameter scan was performed on events containing quarks ($ee \rightarrow qq\nu\nu$) to evaluate the fake rate, where a quark jet could be misidentified as a $\tau$ jet. The final parameters selected for $\tau$ lepton reconstruction at 1.4~TeV are as follows:
\begin{itemize}
    \item Minimum $p_T$ to enter reconstruction: 1~GeV
   \item Minimum $p_T$ for $\tau$ seed: 5~GeV
   \item Maximum invariant mass of the $\tau$ candidate: 2.5~GeV
   \item Opening angle of search cone: 0.07~rad
   \item Opening angle of isolation cone (relative to search cone): 0.3~rad
   \item Maximum energy allowed in isolation cone: 5~GeV
\end{itemize}

In addition, further cuts to the $\tau$ candidate where applied to reduce the fake rate of the tau reconstruction:
\begin{itemize}
    \item $p_T>19$ GeV for $1.4$ TeV and $p_T>35$ GeV for $3$ TeV.
    \item $16^o<|\theta_{\tau}|<38^o$ and $42^o<|\theta_{\tau}|<89^o$ for both energy stages.
\end{itemize}

At 1.4~TeV, we successfully reconstructed 
39.07\% of the $\tau$ leptons in the $h \rightarrow \tau e$ signal sample, 
36.90\% in the $h \rightarrow \tau \mu$ sample, and 32.12\% in the 
$ee \rightarrow \ell\ell^{\prime}\nu\nu$ background.  
The fake rate from the $ee \rightarrow qq\nu\nu$ sample is only 3.42\%.

At 3~TeV, the reconstruction efficiency is 
24.03\% for the $h \rightarrow \tau e$ signal, 23.45\% for 
$h \rightarrow \tau \mu$, and 21.38\% for the 
$ee \rightarrow \ell\ell^\prime\nu\nu$ background, while the fake rate from 
$ee \rightarrow qq\nu\nu$ remains low at 1.51\%.

The $ee \rightarrow qq\nu\nu$ fake rates are inclusive over (q=u,d,s,c,b). To investigate the jet-to-tau misidentification rate, we examined the generator-level ancestry of the particles seeding the fake tau-jet candidates. At both energies, approximately \(55\%\) and \(23\%\) of the fake rate originate from light- and strange-hadron activity, respectively. This is physically expected, as their fragmentation can more readily mimic the narrow, low-track-multiplicity signature of hadronic tau decays. Contributions from direct fragmentation and heavy-flavor hadrons are smaller, with a negligible bottom-hadron component.

A summary of the results obtained in the $\tau$ reconstruction stage can be found on Table (\ref{Tab:tau_finder}).

\begin{table}[tb]
\centering
\begin{tabular}{ |c|c|c|c|c|} 
 \hline
 Energy stage ($\sqrt{s}$) & Eff. $h \rightarrow \tau e$ & Eff. $h \rightarrow \tau \mu$ & Eff. $ee \rightarrow ll^\prime\nu\nu$ & Fake rate $ee \rightarrow qq\nu\nu$ \\ 
\hline
$1.4$ TeV & $39.07\%$ & $36.9\%$  & $32.12\%$ & $3.42\%$ \\ 
\hline
$3$ TeV  & $24.03\%$ & $23.45\%$ & $21.38\%$ & $1.51\%$ \\ 
\hline
\end{tabular}
\caption{Selection efficiencies of the TauFinder algorithm for reconstructing $\tau$-lepton candidates, together with the corresponding inclusive fake rates obtained from the ($ee \rightarrow qq\nu\nu$) sample, where (q=u,d,s,c,b), for the different signal and background channels at $\sqrt{s}=1.4$ and $3$~TeV.}
 %\caption{Table showing the efficiencies and fake rates corresponding to the different decay channels for both energy stages.}
 \label{Tab:tau_finder}
\end{table}

\subsubsection{\texorpdfstring{$h\rightarrow\tau\mu$}{h->tau mu} decay channel}

%Although this channel has the highest branching ratio,
The main challenge for this channel lies in the rapid decay of $\tau$ leptons into leptons and quarks, requiring reconstruction through these decay products. This section describes the steps for reconstructing events with one muon and one $\tau$ lepton from PFOs.

\begin{figure}[tbh]
     \centering
     \begin{subfigure}[b]{0.47\textwidth}
         \centering
         \includegraphics[width=\textwidth]{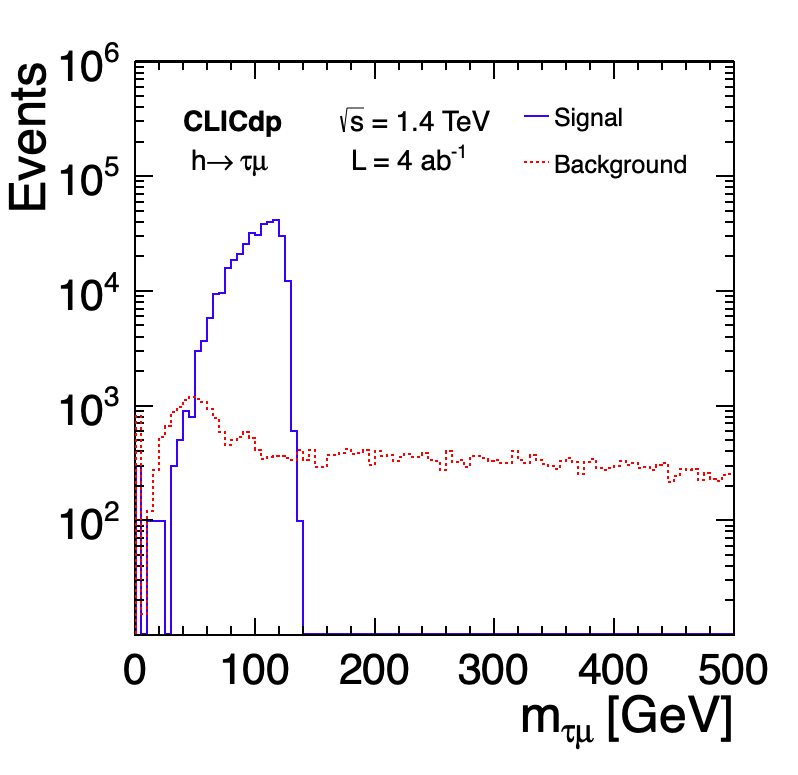}
         \caption{}
         \label{m_taumu}
     \end{subfigure}
     \hfill
     \begin{subfigure}[b]{0.47\textwidth}
         \centering
         \includegraphics[width=\textwidth]{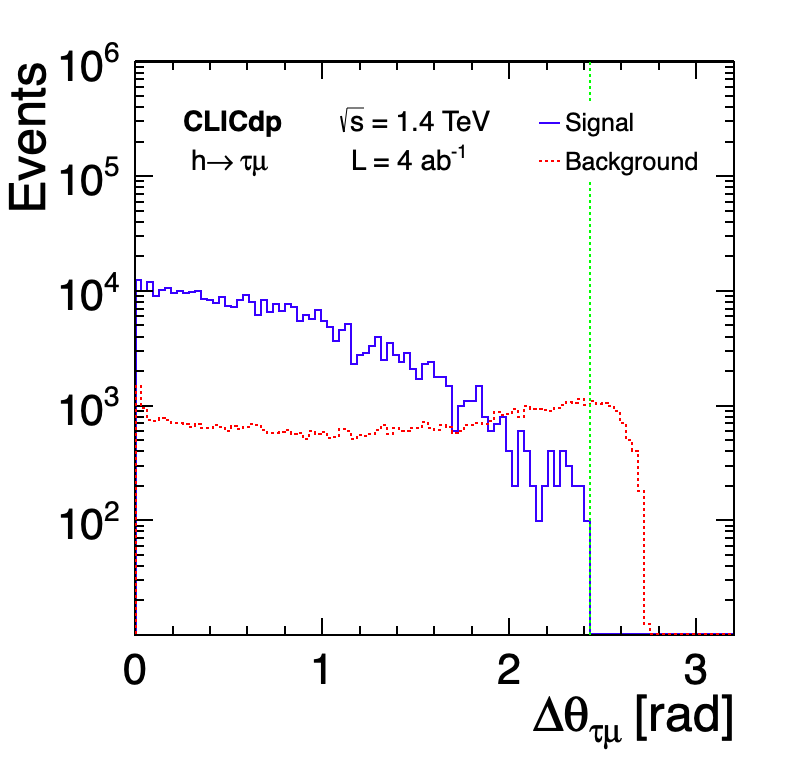}
         \caption{}
         \label{delta_theta_taumu}
     \end{subfigure}
     \hfill
        \caption{(a) Invariant mass distribution of the dilepton system, $m_{\tau\mu}$, and (b) distribution of the polar angle difference between the two leptons, $\Delta\theta_{\tau\mu}$, for events at $\sqrt{s} = 1.4~\text{TeV}$.}
\end{figure}

After $\tau$ lepton reconstruction, the event selection requires two reconstructed, oppositely charged leptons of different flavors ($\tau$, $\mu$). For $\sqrt{s}=1.4$~TeV collisions, a further requirement of  $|\Delta\theta_{\tau \mu}| < 2.43\ \text{rad}$ was imposed. As an example, Figure~\ref{m_taumu} shows the distribution of the invariant mass of the dilepton system for the signal compared to the background considered in this channel. As can be seen, the reconstructed distribution exhibits a peak at the Higgs boson mass, with reduced resolution compared to the previous channels due mainly to the presence of neutrinos in the tau lepton decays. Figure~\ref{delta_theta_taumu} presents the difference in polar angle $\Delta\theta_{\tau \mu}$ at  $1.4~\text{TeV}$, which shows a behaviour similar to the previous channel.

\begin{figure}[tbh]
     \centering
     \begin{subfigure}[b]{0.47\textwidth}
         \centering
         \includegraphics[width=\textwidth]{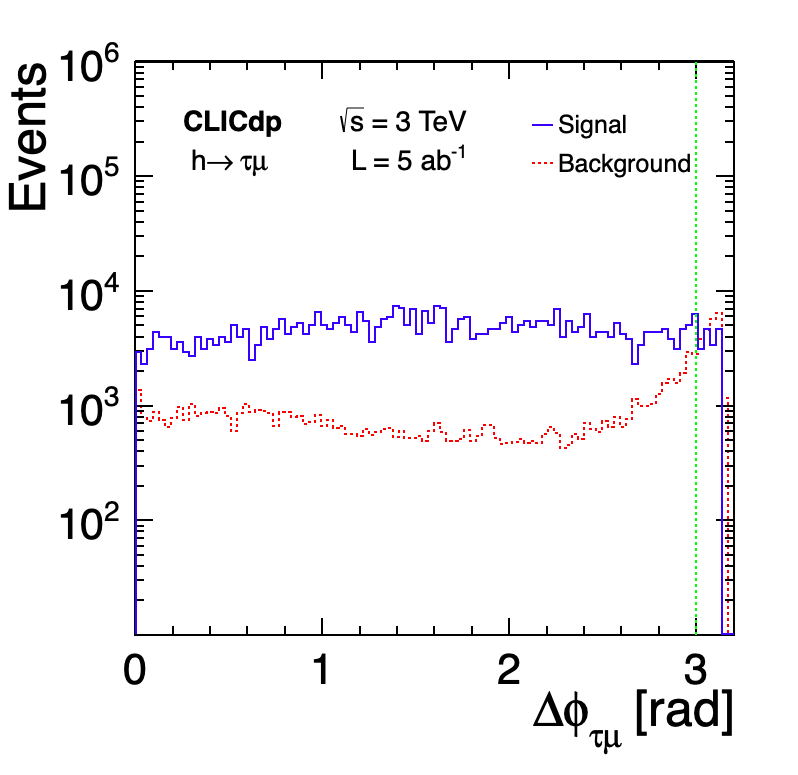}
         \caption{}
         \label{delta_phi_taumu_3TeV}
     \end{subfigure}
     \hfill
     \begin{subfigure}[b]{0.47\textwidth}
         \centering
         \includegraphics[width=\textwidth]{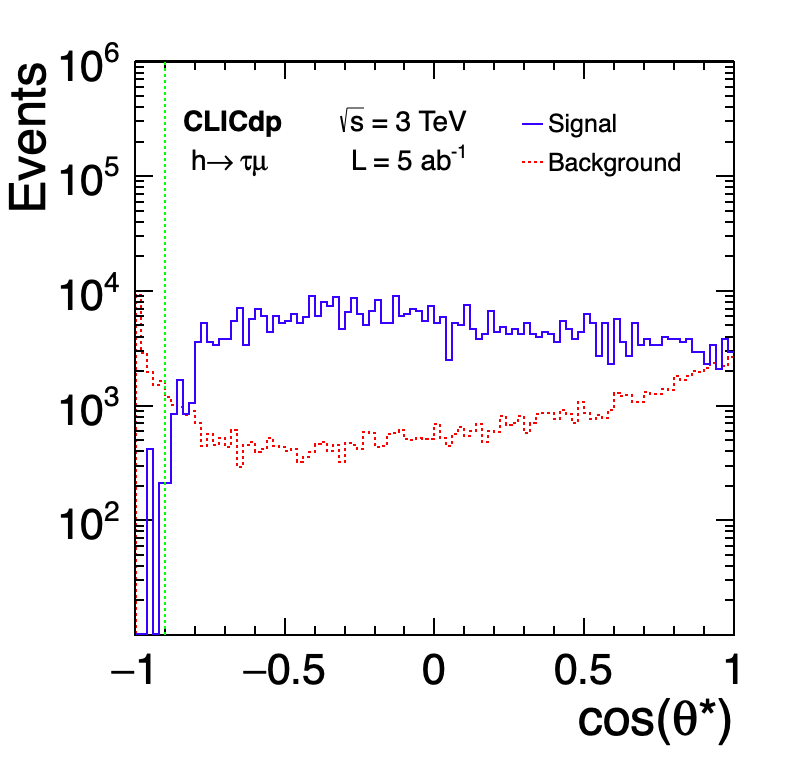}
         \caption{}
         \label{costheta_taumu_3TeV}
     \end{subfigure}
     \hfill
        \caption{(a) The distribution of the azimuthal angle difference between the two leptons, $\Delta\phi_{\tau\mu}$, and (b) the cosine of the helicity angle of the dilepton system $\tau\mu$, for events at $\sqrt{s} = 3~\text{TeV}$.}
\end{figure}

Similarly, for $\sqrt{s}=3~\text{TeV}$, additional requirements of  $|\Delta\phi_{\tau\mu}| < 3.0~\text{rad}$ and $\cos\theta^* > -0.9$ were used, as shown in Figures~\ref{delta_phi_taumu_3TeV} and~\ref{costheta_taumu_3TeV}. The same criteria as before were used to maintain high signal efficiency. At 1.4~TeV, the dominant requirement is on $\Delta\theta_{\tau\mu}$, which removes about 12\% of the background while leaving the signal essentially unaffected. This exploits the fact that background processes tend to produce leptons in nearly back-to-back topologies, whereas signal events from $h\to \tau\mu$ do not. At 3~TeV, the boosted kinematic enhances the role of azimuthal correlations. 
The cut on $\Delta\phi_{\tau\mu}$ rejects nearly a quarter of the background by suppressing planar back-to-back configurations typical of two-fermion and diboson processes, while the helicity angle selection $\cos\theta^*>-0.9$ removes additional extreme backward configurations. After all cuts, 34.9\% of the signal events remain, and the background is reduced to 1.08\% at 1.4~TeV, while 21.1\% of the signal events remain, and the background is reduced to 0.29\% for the 3~TeV sample. An overview of the selection cut results is summarized in Table (\ref{Tab:cutstaumu}) and Table (\ref{Tab:cutstaumu_3TeV}).

\begin{table}[tbh]
\centering
\begin{tabular}{ |c|c|c|c|} 
 \hline
 Process & Events & Efficiency & Exp. Events \\ 
\hline
$ee\rightarrow h\nu\nu\, (h\rightarrow \tau\mu)$ & $9.8$k  & 34.5\% & $338 \pm 6$k\\ 

\hline
$ee\rightarrow \ell\ell^{\prime}\nu\nu$ / $ee\rightarrow qq\nu\nu$  
& $2.0$M & 1.21\% & $61.3 \pm 0.4$k\\ 

\hline
\end{tabular}
 \caption{Number of generated events in signal and background samples at $1.4$ TeV. The rightmost column shows the number of expected events assuming a branching ratio of 1 for signal, $\mathcal{L}=4\,\mathrm{ab}^{-1}$ and the values from Table~\ref{Tab:mcsamples}. The notation is the same as in Table \ref{Tab:cutsemu}.} %Statistical uncertainties reflect the finite size of the simulated samples.}
 \label{Tab:cutstaumu}
\end{table}

\begin{table}[tbh]
\centering
\begin{tabular}{ |c|c|c|c|} 
 \hline
 Process & Events & Efficiency & Exp. Events  \\ 

\hline
$ee\rightarrow h\nu\nu\, (h\rightarrow \tau\mu)$ & $9.9$k & $21.1\%$ & $437 \pm 10$k \\ 

\hline
$ee\rightarrow \ell\ell^{\prime}\nu\nu$ / $ee\rightarrow qq\nu\nu$ & $2.1$M & $0.42\%$ & $29.2 \pm 0.3$k \\ 

\hline
\end{tabular}
 \caption{Number of generated events in signal and background samples at 3~TeV. The rightmost column shows the number of expected events assuming a branching ratio of 1 for signal $\mathcal{L}=5\,\mathrm{ab}^{-1}$ and values from Table~\ref{Tab:mcsamples}. The notation is the same as in Table \ref{Tab:cutsemu}.}%Statistical uncertainties reflect the finite size of the simulated samples.}
 \label{Tab:cutstaumu_3TeV}
\end{table}

The final event selection also employs a BDT classifier \cite{TMVA:2007ngy} utilizing the following kinematic variables. At $1.4$ TeV are $m_{\tau\mu}$, $|\Delta\theta_{\tau\mu}|$, $|\Delta\phi_{\tau\mu}|$, $\cos\theta^*$ and $\theta_{\text{miss}}$; At $3$ TeV the selected variables are $m_{\tau\mu}$, $|\Delta\phi_{\tau\mu}|$, $\cos\theta^*$, and $\theta_{\text{miss}}$. When choosing these variables we checked the correlation matrices and found that no correlation was higher than $54\%$ on both the signal and the background, for the two energy stages. High background rejection and good signal efficiency are obtained for both energy schemes, as indicated by area-under-the-ROC-curve values close to 1.

Using the same statistical approach as the previous section, the expected 95\%~CL limit for the case that no signal is observed at an integrated luminosity of 4~ab$^{-1}$ and $\sqrt{s}=1.4$~TeV is 
$$BR(h\rightarrow \tau\mu)<1.70\times 10^{-4}.$$
The equivalent expected 95\%~CL limit for an integrated  luminosity of 5~ab$^{-1}$ at $\sqrt{s}=3$~TeV is
$$BR(h\rightarrow \tau\mu)< 8.41 \times 10^{-5}.$$

The impact of background normalization uncertainties remains below \(1\%\), with a 5\% uncertainty degrading the expected limit by approximately \(0.95\%\) at 1.4 TeV and \(0.86\%\) at 3 TeV. The corresponding expected \(5\sigma\) discovery reach is \(\mathrm{BR}(H\to\tau\mu)=7.36\times10^{-4}\) at 1.4 TeV and \(4.15\times10^{-4}\) at 3 TeV.

These results demonstrate that CLIC  operation at 1.4~TeV achieves a sensitivity that is 10 times better than that of the CMS and ATLAS experiments, while at 3~TeV this sensitivity is a factor of around 21 times better,  see~\cite{ATLASLFV,CMSLFV}.

\subsubsection{\texorpdfstring{$h\rightarrow\tau e$}{h->tau e} decay channel}

The measurement of the LFV Higgs boson decay into an electron and a $\tau$ in the final state faces similar challenges to those of the previous channel due to the need to reconstruct the $\tau$ lepton from its decay products. The steps for reconstructing events with one electron and one $\tau$ lepton from PFOs are detailed in this section.

\begin{figure}[tbh]
     \centering
     \begin{subfigure}[b]{0.47\textwidth}
         \centering
         \includegraphics[width=\textwidth]{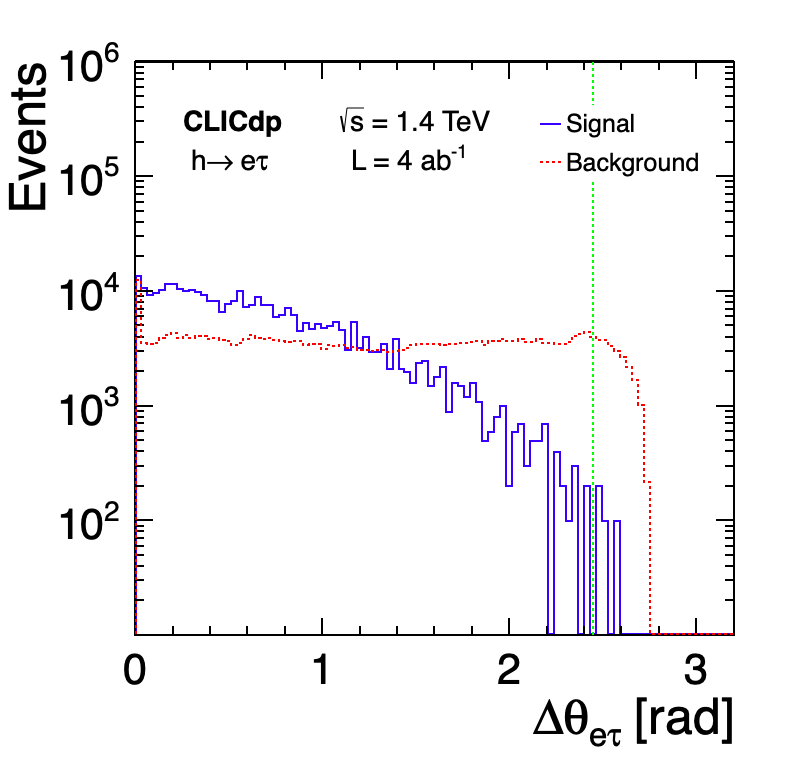}
         \caption{}
         \label{deltatheta_taue}
     \end{subfigure}
     \hfill
     \begin{subfigure}[b]{0.47\textwidth}
         \centering
         \includegraphics[width=\textwidth]{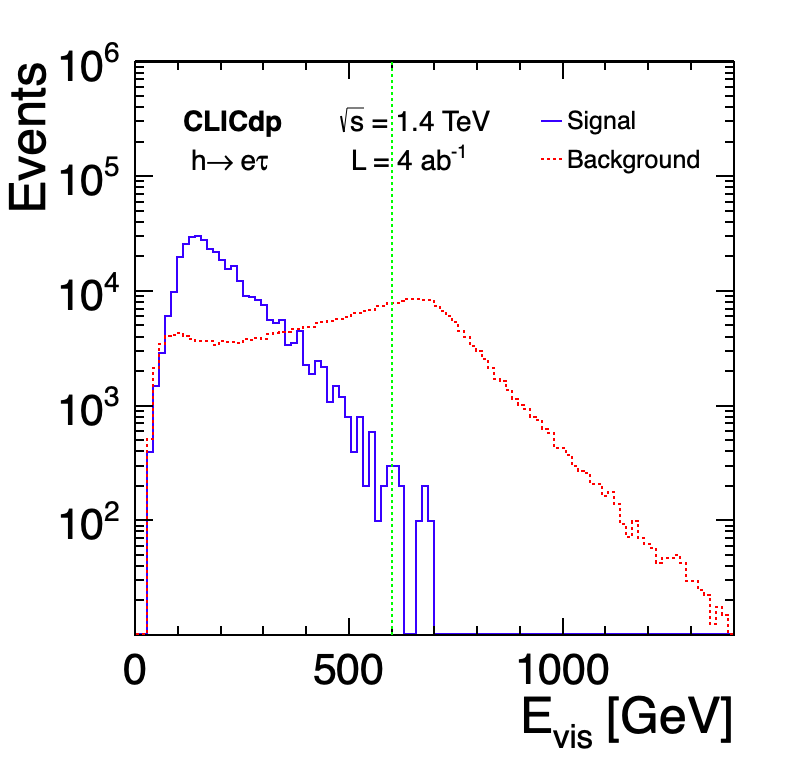}
         \caption{}
         \label{Evis_taue}
     \end{subfigure}
     \hfill
        \caption{(a) Distribution of the polar angle difference between the two leptons, $\Delta\theta_{\tau e}$, and (b) the visible energy distribution of the events at $\sqrt{s}=1.4~\text{TeV}$.
}
\end{figure}

\begin{figure}[tbh]
     \centering
     \begin{subfigure}[b]{0.47\textwidth}
         \centering
         \includegraphics[width=\textwidth]{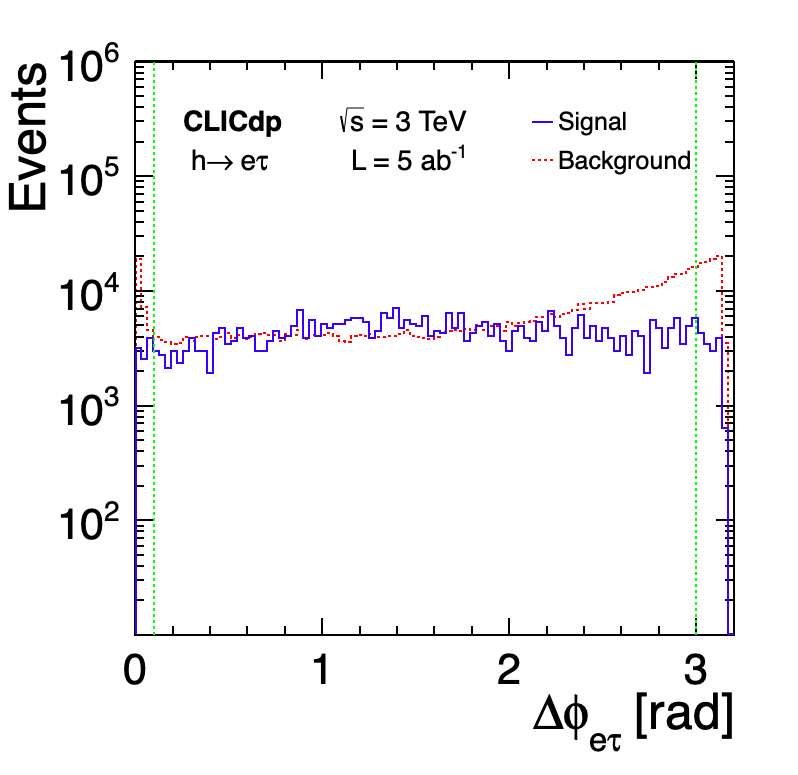}
         \caption{}
         \label{deltatphi_taue_3}
     \end{subfigure}
     \hfill
     \begin{subfigure}[b]{0.48\textwidth}
         \centering
         \includegraphics[width=\textwidth]{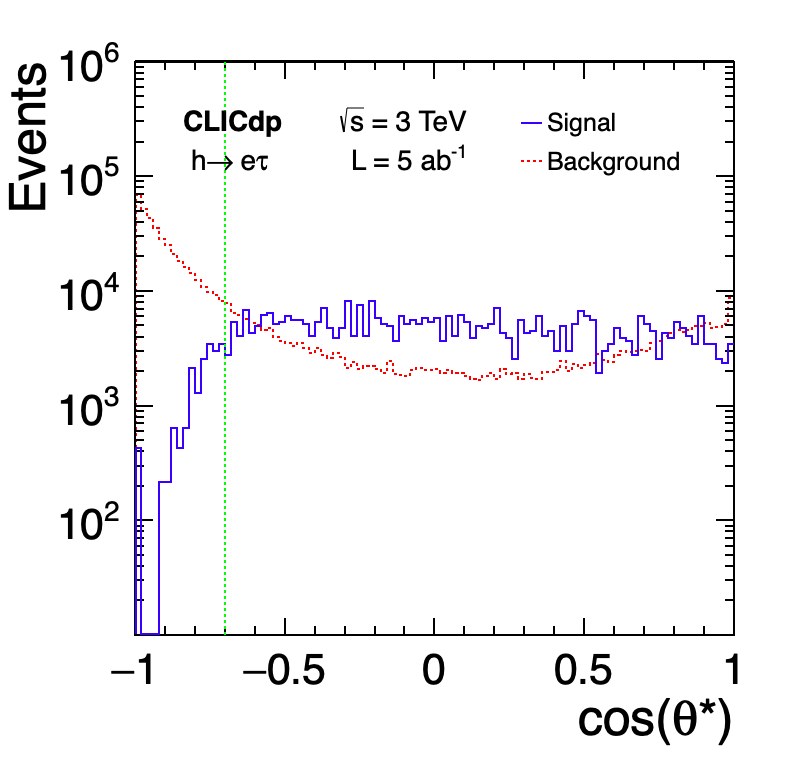}
         \caption{}
         \label{costheta_taue_3}
     \end{subfigure}
     \hfill
        \caption{(a) The distribution of the azimuthal angle difference between the two leptons, $\Delta\phi_{\tau e}$, and (b) the cosine of the helicity angle of the dilepton system $\tau e$, for events at $\sqrt{s} = 3~\text{TeV}$.
}
\end{figure}

The event selection requires two reconstructed, oppositely charged leptons of different flavors ($\tau$, e), where the electron has an energy $E_{\text{e}}>8$ GeV to avoid poorly reconstructed electrons for the same reason as in the $h \rightarrow e \mu$ channel. Bremsstrahlung photons were included in the invariant mass calculation. Additionally, at  1.4~TeV, the cuts $|\Delta\theta_{\tau e}| < 2.45~\text{rad}$ and $E_{\text{vis}} < 600~\text{GeV}$ were applied, as shown in Figures~\ref{deltatheta_taue} and~\ref{Evis_taue}. At 3~TeV, the cuts $0.1~\text{rad} < |\Delta\phi_{\tau e}| < 3.0~\text{rad}$ and $\cos\theta^* > -0.7$ were used, as shown in Figures~\ref{deltatphi_taue_3} and~\ref{costheta_taue_3}. The same criteria as the previous channels were used to maintain high signal efficiency. At 1.4~TeV, the most effective requirement is on $E_{\text{vis}}$, which suppresses nearly 40\% of the background arising from multibody or ISR-enhanced events, while leaving the signal essentially unaffected. At 3~TeV, the boosted kinematics enhance the discriminating power of angular correlations. The $\cos\theta^*$ cut removes over a third of the background by rejecting extreme backward configurations, while $\Delta\phi_{e\tau}$ cut trims back-to-back topologies. After all cuts, 34.7\% of the signal events remain, and the background is reduced to 3.49\% at 1.4~TeV, while 18.4\% of the signal events remain, and the background is reduced to 0.93\% for the 3~TeV sample. An overview of the selection cut results is summarized in Table~\ref{Tab:cutsetau} and Table~\ref{Tab:cutsetau_3TeV}.

\begin{table}[tbh]
\centering
\begin{tabular}{ |c|c|c|c|} 
 \hline
 Process & Events &  Efficiency & Exp. Events \\ 
\hline
$ee\rightarrow h\nu\nu\, (h\rightarrow e\tau)$ & $9.9$k & $34.7\%$ & $338 \pm 6$k \\ 

\hline
$ee\rightarrow \ell\ell^{\prime}\nu\nu$ / $ee\rightarrow qq\nu\nu$  
& $2.0$M  & $4.8\%$ & $244.7 \pm 0.8$k  \\ 

\hline
\end{tabular}
 \caption{Number of generated events in signal and background samples at $1.4$ TeV. The rightmost column shows the number of expected events assuming a branching ratio of 1 for signal, $\mathcal{L}=4\,\mathrm{ab}^{-1}$ and values from Table~\ref{Tab:mcsamples}. The notation is the same as in Table \ref{Tab:cutsemu}.} %Statistical uncertainties reflect the finite size of the simulated samples.}
 \label{Tab:cutsetau}
\end{table}

\begin{table}[tbh]
\centering
\begin{tabular}{ |c|c|c|c|} 
 \hline
 Process & Events & Efficiency & Exp. Events \\ 
\hline
$ee\rightarrow h\nu\nu\, (h\rightarrow e\tau)$ 
& $9.7$k & $20.4\%$& $424 \pm 9$k \\ 

\hline
$ee\rightarrow \ell\ell^{\prime}\nu\nu$ / $ee\rightarrow qq\nu\nu$  
& $2.1$M & $1.13\%$& $77.8 \pm 0.6$k  \\ 

\hline
\end{tabular}
 \caption{Number of generated events in signal and background samples at $3$ TeV. The last column shows the expected events assuming  a branching ratio of 1 for signal, $\mathcal{L}=5\,\mathrm{ab}^{-1}$ and values from Table~\ref{Tab:mcsamples}. The notation is the same as in Table \ref{Tab:cutsemu}.} %Statistical uncertainties reflect finite simulated sample sizes.}
 \label{Tab:cutsetau_3TeV}
\end{table}

The final event selection also employs a BDT classifier \cite{TMVA:2007ngy} utilizing the following kinematic variables At $1.4$ TeV are $m_{\tau e}$, $|\Delta\theta_{\tau e}|$, $|\Delta\phi_{\tau e}|$, $\beta_{\tau e}$, $\cos\theta^*$, $E_{\text{vis}}$, $E_{\text{rest}}$ and $\theta_{\text{miss}}$. At $3$ TeV the selected variables are $m_{\tau e}$, $|\Delta\theta_{\tau e}|$, $|\Delta\phi_{\tau e}|$, $\cos\theta^*$ and $E_{\text{vis}}$. When choosing these variables we checked the correlation matrices and made sure that no correlation was higher than $68\%$ on both the signal and the background, for the two energy stages. High background rejection and good signal efficiency are obtained for both energy schemes, as indicated by area-under-the-ROC-curve values close to 1.

Again, using the same statistical approach as before, the expected 95\% CL limit for the case that no signal is present at an integrated luminosity of 4~ab$^{-1}$ at $\sqrt{s}=1.4$~TeV is given by
$$BR(h\rightarrow \tau e)<1.94\times 10^{-4}.$$
The equivalent expected 95\% CL limit for an integrated luminosity of 5~ab$^{-1}$ at $\sqrt{s}=3$~TeV is
$$BR(h\rightarrow \tau e)< 2.08 \times 10^{-4}.$$
The impact of background normalization uncertainties remains small, with a 5\% uncertainty degrading the expected limit by approximately \(0.34\%\) at 1.4 TeV and \(1.74\%\) at 3 TeV. The corresponding expected \(5\sigma\) discovery reach is \(\mathrm{BR}(H\to e\tau)=1.02\times10^{-3}\) at 1.4 TeV and \(8.08\times10^{-4}\) at 3 TeV.

These results demonstrate that CLIC operation at 1.4~TeV has the potential to reach a sensitivity 10 times greater than the latest results from the ATLAS and CMS experiments \cite{ATLASLFV,CMSLFV}, for both energy stages.

\section{Conclusions}
This paper presents a detailed sensitivity study on LFV Higgs boson decays at the proposed future electron-positron linear collider CLIC, utilizing full detector simulation and event reconstruction. These very rare decay modes provide a unique window into physics beyond the Standard Model, with potential connections to neutrino mass generation, charged lepton flavor violation and new symmetry-breaking mechanisms. Three different final states were explored: $h\rightarrow\mu\tau$, $h\rightarrow e\tau$, and $h\rightarrow e\mu$, assuming CLIC center-of-mass energy stages at $1.4$ TeV and $3$ TeV for integrated luminosities of $\mathcal{L}=4$ ab$^{-1}$ and $\mathcal{L}$=5 ab$^{-1}$, respectively. The expected sensitivities are of order $10^{-5}$ for the $h\rightarrow e\mu$ decay and of order $10^{-4}$--$10^{-5}$ for inclusive $\tau$ channels. These studies assumed unpolarised electron beams. The main limitations are the residual electroweak backgrounds and, in the tau channels, the tau-reconstruction efficiency and missing neutrinos. Further improvement of the sensitivities might be achieved through optimized multivariate and tau-reconstruction techniques, polarized beams that are part of the current CLIC baseline scenario and larger simulated samples. At the same time, future studies are expected to provide a more complete treatment of detector and background-modeling uncertainties.

\section{Acknowledgments}

Francisca Garay acknowledges support from ANID FONDECYT grant No. 3180653 and also acknowledge support from ANID Millennium Science Initiative Program ICN2019 044.

% add references
\printbibliography[title=References]

@Article{ATLASHiggs,
  author        = {{The ATLAS collaboration}},
  title         = "Observation of a new particle in the search for the Standard Model Higgs boson with the ATLAS detector at the LHC",
  journal       = "Phys.Lett. B716 (2012) 1",
  year = "",
  doi           = "https://doi.org/10.1016/j.physletb.2012.08.020",
}

@Article{CMSHiggs,
  author        = {{The CMS collaboration}},
  title         = "Observation of a new boson at a mass of 125 GeV with the CMS experiment at the LHC",
  journal       = "Phys.Lett. B716 (2012) 30",
  doi           = "https://doi.org/10.1016/j.physletb.2012.08.021",
}

@Article{Higgs1,
  author        = {{F. Englert and R. Brout}},
  title         = "Broken Symmetry and the Mass of Gauge Vector Mesons",
  journal       = "Phys.Rev.Lett.13 (1964) 321",
  volume        = "",
  number        = "",
  month         = "",
  year          = "",
  pages         = "",
  doi           = "https://doi.org/10.1103/PhysRevLett.13.321",
  OPTurl           = "",
  note          = "",
}

@Article{Higgs2,
  author        = {{P.W. Higgs}},
  title         = "Broken symmetries, massless particles and gauge fields",
  journal       = "Phys. Lett. 12 (1964) 132",
  volume        = "",
  number        = "",
  month         = "",
  year          = "",
  pages         = "",
  doi           = "https://doi.org/10.1016/0031-9163(64)91136-9",
  OPTurl           = "",
  note          = "",
}

@Article{Higgs5,
  author        = {{G. Guralnik,C. Hagen and T. Kibble}},
  title         = "Global Conservation Laws and Massless Particles",
  journal       = "Phys.Rev.Lett.13 (1964) 585",
  volume        = "",
  number        = "",
  month         = "",
  year          = "",
  pages         = "",
  doi           = "https://doi.org/10.1103/PhysRevLett.13.585",
  OPTurl           = "",
  note          = "",
}

@Article{LFVTheo1,
  author        = {{J.D.Bjorken and S. Weinberg}},
  title         = "Mechanism for Non conservation of Muon Number",
  journal       = "Phys.Rev.Lett.38 (1977) 622",
  volume        = "",
  number        = "",
  month         = "",
  year          = "",
  pages         = "",
  doi           = "https://doi.org/10.1103/PhysRevLett.38.622",
  OPTurl           = "",
  note          = "",
}

@Article{LFVTheo2,
  author        = {{J.L. Diaz-CruzandJ.J.Toscano}},
  title         = "Lepton flavor violating decays of Higgs bosons beyond the standard model",
  journal       = "Phys.Rev.D 62 (2000) 116005",
  volume        = "",
  number        = "",
  month         = "",
  year          = "",
  pages         = "",
  doi           = "https://doi.org/10.1103/PhysRevD.62.116005",
  OPTurl           = "",
  note          = "",
}

@Article{LFVTheo3,
  author        = {{M.Arana-Catania, E. Arganda and M. J. Herrero}},
  title         = "Non-decoupling SUSY in LFV Higgs decays: a window to new physics at the LHC",
  journal       = "JHEP09 (2013) 160, (Erratum:JHEP 10 (2015) 192)",
  volume        = "",
  number        = "",
  month         = "",
  year          = "",
  pages         = "",
  doi           = "https://doi.org/10.1007/JHEP09(2013)160",
  OPTurl           = "",
  note          = "",
}

@Article{LFVTheo4,
  author        = {{A. Arhrib, Y. Cheng and O. C. W. Kong}},
  title         = "Comprehensive analysis on lepton flavor violating Higgs
boson to $\mu\tau$ decay insupersymmetry without R parity",
  journal       = "JPhys. Rev.D 87 (2013) 015025",
  volume        = "",
  number        = "",
  month         = "",
  year          = "",
  pages         = "",
  doi           = "https://doi.org/10.1103/PhysRevD.87.015025",
  OPTurl           = "",
  note          = "",
}

@Article{LFVTheo5,
  author        = {{D. Aloni,Y. Nir and E. Stamou}},
  title         = "Large $BR(h\rightarrow \tau\mu)$ in the MSSM?",
  journal       = "JHEP04 (2016) 162",
  volume        = "",
  number        = "",
  month         = "",
  year          = "",
  pages         = "",
  doi           = "https://doi.org/10.1007/JHEP04(2016)162",
  OPTurl           = "",
  note          = "",
}

@Article{LFVTheo6,
  author        = {{K. Agashe and R. Contino}},
  title         = "Composite Higgs-mediated flavor-changing neutral current",
  journal       = "Phys.Rev.D 80 (2009) 075016",
  volume        = "",
  number        = "",
  month         = "",
  year          = "",
  pages         = "",
  doi           = "https://doi.org/10.1103/PhysRevD.80.075016",
  OPTurl           = "",
  note          = "",
}

@Article{LFVTheo7,
  author        = {{A. Azatov, M. Toharia and L. Zhu}},
  title         = "Higgs mediated flavor changing neutral currents in warped extra dimensions",
  journal       = "Phys.Rev.D 80 (2009) 035016",
  volume        = "",
  number        = "",
  month         = "",
  year          = "",
  pages         = "",
  doi           = "https://doi.org/10.1103/PhysRevD.80.035016",
  OPTurl           = "",
  note          = "",
}

@Article{LFVTheo8,
  author        = {{H. Ishimori et al}},
  title         = "Non-Abelian Discrete Symmetries in Particle Physics",
  journal       = "Prog.Theor.Phys. Suppl.183 (2010) 1",
  volume        = "",
  number        = "",
  month         = "",
  year          = "",
  pages         = "",
  doi           = "https://doi.org/10.1143/PTPS.183.1",
  OPTurl           = "",
  note          = "",
}

@Article{LFVTheo9,
  author        = {{G. Perez and L. Randall}},
  title         = "Natural neutrino masses and mixings from warped geometry",
  journal       = "JHEP01 (2009) 077",
  volume        = "",
  number        = "",
  month         = "",
  year          = "",
  pages         = "",
  doi           = "https://doi.org/10.1088/1126-6708/2009/01/077",
  OPTurl           = "",
  note          = "",
}

@Article{LFVTheo10,
  author        = {{M. E. Albrecht et al}},
  title         = "Electroweak and Flavour Structure of a Warped Extra Dimension with Custodial Protection",
  journal       = "JHEP 0909:064 (2009)",
  volume        = "",
  number        = "",
  month         = "",
  year          = "",
  pages         = "",
  doi           = "https://doi.org/10.1088/1126-6708/2009/09/064",
  OPTurl           = "",
  note          = "",
}

@Article{LFVTheo11,
  author        = {{A. Goudelis, O. Lebedev and J.h. Park}},
  title         = "Higgs-induced lepton flavor violation",
  journal       = "Phys.Lett. B 707(2012)369",
  volume        = "",
  number        = "",
  month         = "",
  year          = "",
  pages         = "",
  doi           = "https://doi.org/10.1016/j.physletb.2011.12.059",
  OPTurl           = "",
  note          = "",
}

@Article{LFVTheo12,
  author        = {{D. McKeen, M.Pospelov and A.Ritz}},
  title         = "Modified Higgs branching ratios versus CP and lepton flavor violation",
  journal       = "Phys.Rev.D 86 (2012) 113004",
  volume        = "",
  number        = "",
  month         = "",
  year          = "",
  pages         = "",
  doi           = "https://doi.org/10.1103/PhysRevD.86.113004",
  OPTurl           = "",
  note          = "",
}

@Article{Babar,
  author        = {{Babar Collaboration}},
  title         = "Evidence for an Excess of $\bar{B}\rightarrow D^{(*)}\tau^- \bar{\nu_{\tau}}$ Decays",
  journal       = "Phys.Rev.Lett.109 (2012) 101802",
  volume        = "",
  number        = "",
  month         = "",
  year          = "",
  pages         = "",
  doi           = "https://doi.org/10.1103/PhysRevLett.109.101802",
  OPTurl           = "",
  note          = "",
}

@Article{LHCb1,
  author        = {{LHCb Collaboration}},
  title         = "Measurement of the ratio of the $B^0\rightarrow D^{*-}\tau^+ \nu_{\tau}$ and $B^0\rightarrow D^{*-}\mu^+ \nu_{\mu}$ branching fractions using three-prong $\tau$-lepton decays",
  journal       = "Phys. Rev. Lett. 120, 171802 (2018)",
  volume        = "",
  number        = "",
  month         = "",
  year          = "",
  pages         = "",
  doi           = "https://doi.org/10.1103/PhysRevLett.120.171802",
  OPTurl           = "",
  note          = "",
}

@Article{LHCb2,
  author        = {{LHCb Collaboration}},
  title         = "Measurement of the Ratio of Branching Fraction $\mathcal{B}(\bar{B}^0\rightarrow D^{*+}\tau^- \bar{\nu_{\tau}})/\mathcal{B}(\bar{B}^0\rightarrow D^{*+}\mu^- \bar{\nu_{\mu}})$",
  journal       = "Phys. Rev. Lett. 115, 111803 (2015)",
  volume        = "",
  number        = "",
  month         = "",
  year          = "",
  pages         = "",
  doi           = "https://doi.org/10.1103/PhysRevLett.131.111802",
  OPTurl           = "",
  note          = "",
}

@article{LHCbRK_PRL2023,
  author = {{LHCb Collaboration}},
  title = {Tests of lepton universality using $B^0 \to K_s^0 \ell^+ \ell^-$ and $B^+ \to K^{*+} \ell^+ \ell^-$ decays},
  journal = {Phys. Rev. Lett.},
  volume = {128},
  year = {2023},
  pages = {191802},
  doi = {https://doi.org/10.1103/PhysRevLett.128.191802}
}

@article{LHCbRKstar_PRD2023,
  author = {{LHCb Collaboration}},
  title = {Measurement of the branching fraction ratio $R_K$ at large dilepton invariant mass},
  journal = {J. High Energ. Phys. 2025, 198 (2025)},
  volume = {},
  year = {},
  pages = {},
  doi = {https://doi.org/10.1007/JHEP07(2025)198}
}

@Article{pheno1,
  author        = {{P. Ko,Y. Omura, Y. Shigekami and C. Yu}},
  title         = "LHCb anomaly and B physics in flavored $Z'$ models with flavored Higgs doublets",
  journal       = "Phys.Rev.D 95(2017)115040",
  volume        = "",
  number        = "",
  month         = "",
  year          = "",
  pages         = "",
  doi           = "https://doi.org/10.1103/PhysRevD.95.115040",
  OPTurl           = "",
  note          = "",
}

@Article{pheno2,
  author        = {{A. Carmona and F. Goertz}},
  title         = "Lepton Flavor and Non-Universality from Minimal Composite Higgs Setups",
  journal       = "Phys.Rev.Lett.116 (2016) 251801",
  volume        = "",
  number        = "",
  month         = "",
  year          = "",
  pages         = "",
  doi           = "https://doi.org/10.1103/PhysRevLett.116.251801",
  OPTurl           = "",
  note          = "",
}

@Article{pheno3,
  author        = {{F. J. Botella, G. C. Branco, M. Nebot and M. N. Rebelo}},
  title         = "Flavour-changing Higgs couplings in a class of two Higgs doublet models",
  journal       = "Eur.Phys. J.C 76(2016)161",
  volume        = "",
  number        = "",
  month         = "",
  year          = "",
  pages         = "",
  doi           = "https://doi.org/10.1140/epjc/s10052-016-3993-0",
  OPTurl           = "",
  note          = "",
}

@Article{ATLASLFV,
  author        = {{The ATLAS collaboration}},
  title         = "Searches for lepton-flavour-violating decays of the Higgs boson into $e\tau$ and $\mu\tau$ in $\sqrt{s} = 13$ TeV pp collisions with the ATLAS detector",
  journal       = "JHEP 07 (2023) 166",
  volume        = "",
  number        = "",
  month         = "",
  year          = "",
  pages         = "",
  doi           = "https://doi.org/10.1007/JHEP07%282023%29166",
  OPTurl           = "",
  note          = "",
}

@Article{CMSLFV,
  author        = {{The CMS collaboration}},
  title         = "Search for lepton-flavor violating decays of the Higgs boson in the $\mu\tau$ and $e\tau$ final states in proton-proton collisions at $\sqrt{s} = 13$ TeV",
  journal       = "Phys. Rev. D 104, 032013 (2021)",
  volume        = "",
  number        = "",
  month         = "",
  year          = "",
  pages         = "",
  doi           = "https://doi.org/10.1103/PhysRevD.104.032013",
  OPTurl           = "https://doi.org/10.1103/PhysRevD.104.032013",
  note          = "",
}

@Article{ATLASLFV2,
  author        = {{The ATLAS collaboration}},
  title         = "Search for the Higgs boson decays $h\rightarrow \mu\mu$ and $h\rightarrow e\mu$ in $pp$ collisions
at $\sqrt{s}=13$ TeV with the ATLAS detector",
  journal       = "Phys. Lett.B801 (2020) 135148",
  volume        = "",
  number        = "",
  month         = "",
  year          = "",
  pages         = "",
  doi           = "https://doi.org/10.1016/j.physletb.2019.135148",
  OPTurl           = "",
  note          = "",
}

@Article{Futureee,
  author  = {Qin qin, et al.},
  title   = {Charged lepton flavour violating Higgs decays at future $e^+e^-$ colliders},
  journal = {arXiv:1711.07243v2 [hep-ph]},
  year    = {2018},
  eprint  = {},
}

@Article{FCCeeLFV,
  author  = {P. Sriling and N. Srimanobhas and P. Uttayarat and R. Uttho and V. Wachirapusitanand},
  title   = {Searching for Lepton Flavor Violating decays of the Higgs Boson into $\mu\tau$, $e\tau$, and $e\mu$ final states at FCC-ee},
  journal = {arXiv:2605.30971 [hep-ph]},
  year    = {2026},
  eprint  = {2605.30971},
}

@Article{pheno4,
  author        = {{J.Adam et al.}},
  title         = "New Constraint on the Existence of the $\mu\rightarrow e\gamma$ Decay",
  journal       = "Phys.Rev.Lett.110 (2013) 201801",
  volume        = "",
  number        = "",
  month         = "",
  year          = "",
  pages         = "",
  doi           = "https://doi.org/10.1103/PhysRevLett.110.201801",
  OPTurl           = "",
  note          = "",
}

@Article{pheno5,
  author        = {{R. Harnik, J.Kopp and J. Zupan}},
  title         = "Flavor violating Higgs decays",
  journal       = "JHEP 03(2013)026",
  volume        = "",
  number        = "",
  month         = "",
  year          = "",
  pages         = "",
  doi           = "https://doi.org/10.1007/JHEP03%282013%29026",
  OPTurl           = "",
  note          = "",
}

@Article{pheno6,
  author        = {{G. Blankenburg, J. Ellis and G. Isidori}},
  title         = "Flavour-changing decays of a 125 GeV Higgs-like particle",
  journal       = "Phys. Lett.B 712 (2012) 386",
  volume        = "",
  number        = "",
  month         = "",
  year          = "",
  pages         = "",
  doi           = "https://doi.org/10.1016/j.physletb.2012.05.007",
  OPTurl           = "",
  note          = "",
}

@Article{CLICHIGGS,
  author        = {{The CLICdp Collaboration}},
  title         = "Higgs Physics at the CLIC Electron-Positron Linear Collider",
  journal       = "European Physics Journal",
  volume        = "77",
  number        = "",
  month         = Feb,
  year          = "2017",
  pages         = "475",
  doi           = "https://doi.org/10.1140/epjc/s10052-017-4968-5",
  OPTurl           = "",
  note          = "",
}

@Article{Whiz,
  author        = {{W. Kilian, T. Ohl, J. Reuter}},
  title         = "WHIZARD: simulating multi-particle processes at LHC and ILC",
  journal       = "Eur.Phys.J.C71:1742,2011",
  volume        = "",
  number        = "",
  month         = "",
  year          = "",
  pages         = "",
  doi           = "https://doi.org/10.1140/epjc/s10052-011-1742-y",
  OPTurl           = "",
  note          = "",
}

@Article{CLIC_1,
  author        = {{The CLIC and CLICdp Collaborations}},
  title         = "Updated baseline for a staged Compact Linear Collider",
  journal       = "CERN-2016-004",
  volume        = "",
  number        = "",
  month         = "",
  year          = "2016",
  pages         = "",
  doi           = "https://doi.org/10.5170/CERN-2016-004",
  OPTurl           = "",
  note          = "",
}

@Article{CLIC_2,
  author        = {{The CLIC and CLICdp Collaborations}},
  title         = "The Compact Linear Collider (CLIC) - 2018 Summary Report",
  journal       = "CERN-2018-005-M",
  volume        = "",
  number        = "",
  month         = "",
  year          = "2018",
  pages         = "",
  doi           = "https://doi.org/10.23731/CYRM-2018-002",
  OPTurl           = "",
  note          = "",
}

@Article{CLIC_3,
  author        = {{A. Robson, P. Roloff}},
  title         = "Updated CLIC luminosity staging baseline and Higgs coupling prospects",
  journal       = "CLICdp-Note-2018-002",
  volume        = "",
  number        = "",
  month         = "",
  year          = "2018",
  pages         = "",
  doi           = "https://doi.org/10.48550/arXiv.1812.01644",
  OPTurl           = "",
  note          = "",
}

@Article{CLIC_4,
  author        = {{M. Aicheler et al.}},
  title         = "A Multi-TeV Linear Collider based on CLIC Technology: CLIC Conceptual Design Report",
  journal       = "CERN-2012-007",
  volume        = "",
  number        = "",
  month         = "",
  year          = "2012",
  pages         = "",
  url           = "https://cds.cern.ch/record/1500095",
  OPTurl           = "",
  note          = "",
}

@Article{ILC_2,
  author        = {{H. Abramowicz et al.}},
  title         = "The International Linear Collider Technical Design Report - Volume 4: Detectors",
  journal       = "ILC-REPORT-2013-040",
  volume        = "",
  number        = "",
  month         = "",
  year          = "2013",
  pages         = "",
  doi           = "https://doi.org/10.48550/arXiv.1306.6329",
  OPTurl           = "",
  note          = "",
}

@Article{CLIC_5,
  author        = {{J. Marshall, A. M\"{u}nnich, M. Thomson}},
  title         = "Performance of Particle Flow Calorimetry at CLIC",
journal = {Nuclear Instruments and Methods in Physics Research Section A: Accelerators, Spectrometers, Detectors and Associated Equipment},
volume = {700},
pages = {153-162},
year = {2013},
doi = {https://doi.org/10.1016/j.nima.2012.10.038},
  OPTurl           = "",
  note          = "",
}

@article{CLIC_6,
title = {Particle flow calorimetry and the PandoraPFA algorithm},
journal = {Nuclear Instruments and Methods in Physics Research Section A: Accelerators, Spectrometers, Detectors and Associated Equipment},
volume = {611},
number = {1},
pages = {25-40},
year = {2009},
doi = {https://doi.org/10.1016/j.nima.2009.09.009},
author = {M.A. Thomson},
}

@Article{CLIC_7,
  author        = {{J. S. Marshall, M. A. Thomson}},
  title         = "The Pandora Software Development Kit for Pattern Recognition",
  journal       = "Eur. Phys. J. C 75, 439 (2015)",
  volume        = "",
  number        = "",
  month         = "",
  year          = "",
  pages         = "",
  doi           = "https://doi.org/10.1140/epjc/s10052-015-3659-3",
  OPTurl           = "",
  note          = "",
}

@Article{taufinder,
  author        = {{A. Muennich}},
  title         = "TauFinder: A Reconstruction Algorithm for $\tau$ Leptons at Linear Colliders",
  journal       = "LCD-Note-2010-009",
  volume        = "",
  number        = "",
  month         = "",
  year          = "2010",
  pages         = "",
  doi           = "",
  url           = " http://cds.cern.ch/record/1443551",
  note          = "",
}

@Article{stats,
  author        = {{G. Cowan et al.}},
  title         = "Asymptotic formulae for likelihood-based tests of new physics",
  journal       = "Eur.Phys.J.C71:1554,2011",
  volume        = "",
  number        = "",
  month         = "",
  year          = "",
  pages         = "",
  doi           = "https://doi.org/10.1140/epjc/s10052-011-1554-0",
  OPTurl           = "",
  note          = "",
}

@article{Skrzypek:1990qs,
    author = {{Skrzypek, Maciej and Jadach, Stanislaw}},
    title = "{Exact and approximate solutions for the electron nonsinglet structure function in QED}",
    reportNumber = "TPJU-4-90",
    doi = "https://doi.org/10.1007/BF01483573",
    journal = "Z. Phys. C",
    volume = "49",
    pages = "577--584",
    year = "1991"
}

@article{Sjostrand:2006za,
    author = {{Sjostrand, Torbjorn and Mrenna, Stephen and Skands, Peter Z.}},
    title = "{PYTHIA 6.4 Physics and Manual}",
    doi = "https://doi.org/10.1088/1126-6708/2006/05/026",
    journal = "JHEP 05 (2006) 026",
    volume = "",
    pages = "",
    year = ""
}

@article{OPAL:1995cgp,
    author = {{Alexander, G. and others}},
    collaboration = "OPAL",
    title = "{A Comparison of $b$ and $uds$ quark jets to gluon jets}",
    reportNumber = "CERN-PPE-95-126",
    doi = "https://doi.org/10.1007/BF02907439",
    journal = "Phys. C - Particles and Fields 69 (1995) 543",
    volume = "",
    pages = "",
    year = ""
}

@article{Linssen:2012hp,
    author = {{Linssen et al.}},
    title = "{Physics and Detectors at CLIC: CLIC Conceptual Design Report}",
journal = "",
    eprint = "1202.5940",
    archivePrefix = "arXiv",
    primaryClass = "physics.ins-det",
    reportNumber = "CERN-2012-003, ANL-HEP-TR-12-01, DESY-12-008, KEK-REPORT-2011-7",
    doi = "https://doi.org/10.48550/arXiv.1202.5940",
    month = "2",
    year = "2012"
}

@article{CLICdet_2,
    author = {{Arominski et al.}},
    title = "{A detector for CLIC: main parameters and performance}",
    journal = "",
    primaryClass = "physics.ins-det",
    reportNumber = "CLICdp-Note-2018-005",
    doi = "https://doi.org/10.48550/arXiv.1812.07337",
    month = "",
    year = "2018"
}

@article{Was:2000st,
    author = {{Was, Z.}},
    title = "{TAUOLA the library for tau lepton decay, and KKMC / KORALB / KORALZ /... status report}",
    doi = "https://doi.org/10.1016/S0920-5632%2801%2901200-2",
    journal = "Nucl.Phys.Proc.Suppl. 98 (2001) 96-102",
    volume = "",
    pages = "",
    year = ""
}

@article{Robson:2025,
    author = {{Robson, Aidan and Roloff, Philipp}},
    title = "{CLIC Higgs coupling prospects with 100 Hz operation}",
    journal = "",
    eprint = "arXiv:2503.21857v1",
    archivePrefix = "arXiv",
    primaryClass = "hep-ex",
    reportNumber = "CLICdp-Note-2025-002",
    month = "12",
    year = "2025"
}

@article{Gaede:2006pj,
    author = "Gaede, F.",
    editor = "Blumlein, J. and Friebel, W. and Naumann, T. and Riemann, T. and Wegner, P. and Perret-Gallix, D.",
    title = "{Marlin and LCCD: Software tools for the ILC}",
    doi = "10.1016/j.nima.2005.11.138",
    journal = "Nucl. Instrum. Meth. A",
    volume = "559",
    pages = "177--180",
    year = "2006"
}

@article{TMVA:2007ngy,
    author = "Hoecker, Andreas and others",
    collaboration = "TMVA",
    title = "{TMVA - Toolkit for Multivariate Data Analysis}",
    eprint = "physics/0703039",
    archivePrefix = "arXiv",
    reportNumber = "CERN-OPEN-2007-007",
    month = "3",
    year = "2007"
}

\end{document}